\documentclass[]{aastex631}

\newcommand{\loggcms}{$\log{(\rm{g}/\cmss)}$}
\newcommand{\cmss}{\mbox{$\mbox{cm}\,\mbox{s}^{-2}$}}

\accepted{February 23, 2023}

\shorttitle{NGC 6819 Variable Blue Stragglers}
\shortauthors{Guzik et al.}

\begin{document}

\title{Variable Blue Straggler Stars in Open Cluster NGC 6819 Observed in the {\textit{\textbf{Kepler}}} `Superstamp' Field }

\correspondingauthor{Joyce A. Guzik}
\email{joy@lanl.gov}

\author[0000-0003-1291-1533]{Joyce A. Guzik}
\affiliation{Los Alamos National Laboratory, Los Alamos, NM 87545,  USA}

\author[0000-0001-6446-6617]{Andrzej S. Baran}
\affiliation{ARDASTELLA Research Group}
\affiliation{Astronomical Observatory, University of Warsaw, Al. Ujazdowskie 4, 00-478 Warszawa, Poland}

\author[0000-0001-9907-690X]{Sachu Sanjayan}
\affiliation{ARDASTELLA Research Group}
\affiliation{Centrum Astronomiczne im. Mikołaja Kopernika, Warsaw, Poland}

\author[0000-0003-0963-0239]{P\'eter N\'emeth}
\affiliation{Astronomical Institute of the Czech Academy of Sciences, Czech Republic}
\affiliation{Astroserver.org, F\H{o} t\'er 1, 8533 Malomsok, Hungary}

\author[0000-0001-7017-678X]{Anne M. Hedlund}
\affiliation{Department of Astronomy, New Mexico State University, Las Cruces, NM 88003, USA}
\affiliation{Los Alamos National Laboratory, Los Alamos, NM 87545, USA}

\author[0000-0001-9659-7486]{Jason Jackiewicz}
\affiliation{Department of Astronomy, New Mexico State University, Las Cruces, NM 88003, USA}

\author[0000-0002-2479-4867]{Lori R. Dauelsberg}
\affiliation{Los Alamos National Laboratory, Los Alamos, NM 87545, USA}



\begin{abstract}

NGC 6819 is an open cluster of age 2.4 Gyr that was in the NASA {\it Kepler} spacecraft field of view from 2009 to 2013. The central part of the cluster was observed in a 200 x 200 pixel `superstamp' during these four years in 30-minute cadence photometry, providing a unique long time-series high-precision data set.  The cluster contains ‘blue straggler’ stars, i.e., stars on the main sequence above the cluster turnoff that should have left the main sequence to become red giants.  We present light curves and pulsation frequency analyses derived from custom photometric reductions for five confirmed cluster members--four blue stragglers and one star near the main-sequence turnoff. Two of these stars show a rich spectrum of $\delta$ Scuti pulsation modes, with 236 and 124 significant frequencies identified, respectively, while two stars show mainly low-frequency modes, characteristic of $\gamma$ Doradus variable stars. The fifth star, a known active x-ray binary, shows only several harmonics of two main frequencies. For the two $\delta$ Scuti stars, we use a frequency separation--mean-density relation to estimate mean density, and then use this value along with effective temperature to derive stellar mass and radius.  For the two stars showing low frequencies, we searched for period-spacing sequences that may be representative of gravity-mode or Rossby-mode sequences, but found no clear sequences. The common age for the cluster members, considered along with the frequencies, will provide valuable constraints for asteroseismic analyses, and may shed light on the origin of the blue stragglers.


\end{abstract}



\section{Introduction} \label{sec:intro}

\begin{figure*}
\gridline{\fig{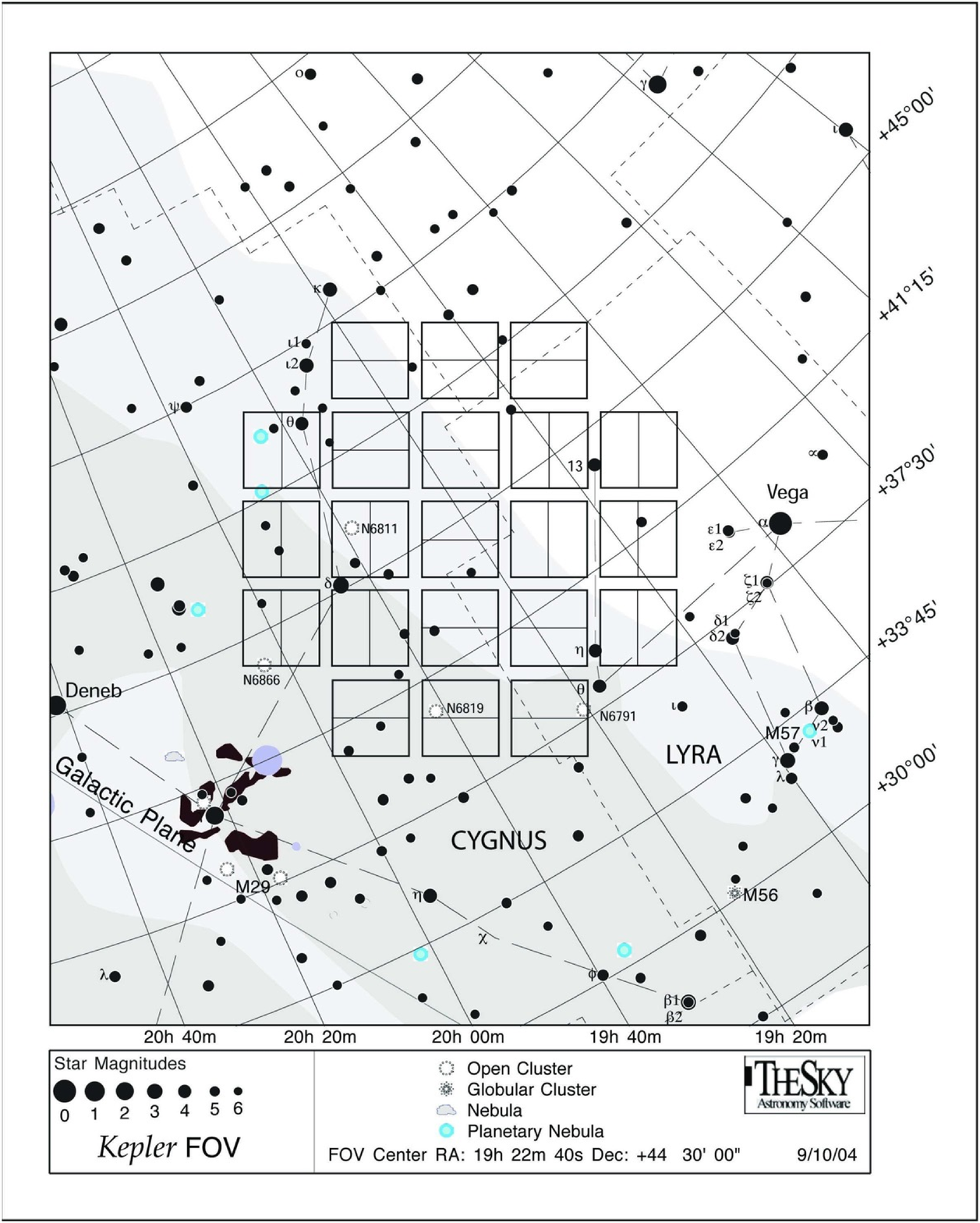}{0.40\textwidth}{(a)}
          \fig{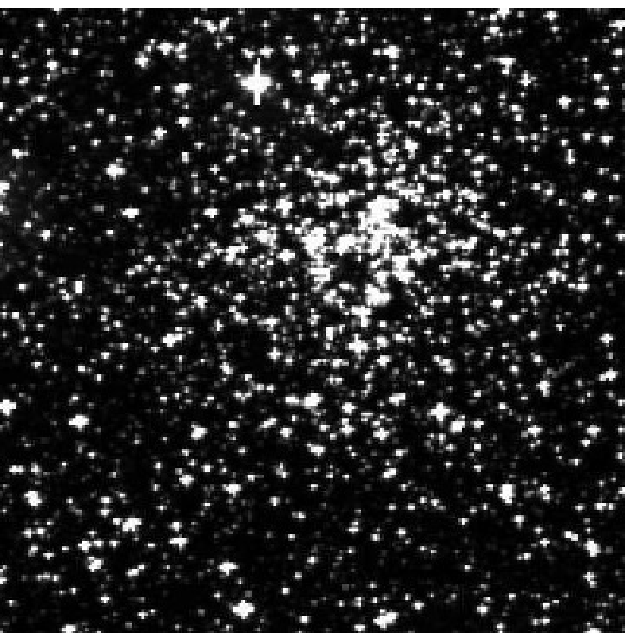}{0.47\textwidth}{(b)}
          }
\caption{(a) Zoom-in on {\it Kepler} original mission field of view showing location of NGC 6819 in the lower center CCD (https://commons.wikimedia.org/wiki/ File: Kepler\_FOV\_hiRes.jpg, NASA/Ames/JPL-Caltech, Image credit: Software Bisque, Public Domain); (b) 200 $\times$ 200 pixel {\it Kepler} superstamp image of the center of NGC 6819 \citep[][reproduced with permission]{2015EPJWC.10106040K}.  {\it Kepler} pixel sizes are 3.98 arsec per pixel.}
\label{fig:Keplerview}
\end{figure*}

NGC 6819 is an open star cluster in the constellation Cygnus discovered by Caroline Herschel in 1784.\footnote{https://en.wikipedia.org/wiki/NGC\_6819} NGC 6819 is about 2.4 billion years old, half the age of the Sun, and around 8000 light years away \citep{2011ApJ...729L..10B, 2013MNRAS.430.3472B, 2016AJ....151...66B}. This cluster was in the NASA {\it Kepler} spacecraft \citep{2010Sci...327..977B, 2010PASP..122..131G} continuous field of view from 2009-2013  (Fig.\,\ref{fig:Keplerview}, left).  The central part of the cluster (Fig.\,\ref{fig:Keplerview}, right) was observed during these four years in 30-minute cadence photometry, providing a unique long time-series high-precision data set for asteroseismology \citep{2015EPJWC.10106040K}. Studying clusters is advantageous for asteroseismology because the cluster members formed together, providing additional modeling constraints such as a common age and element abundances.

\begin{figure}[ht!]
\centering
\includegraphics[width=0.6\textwidth]{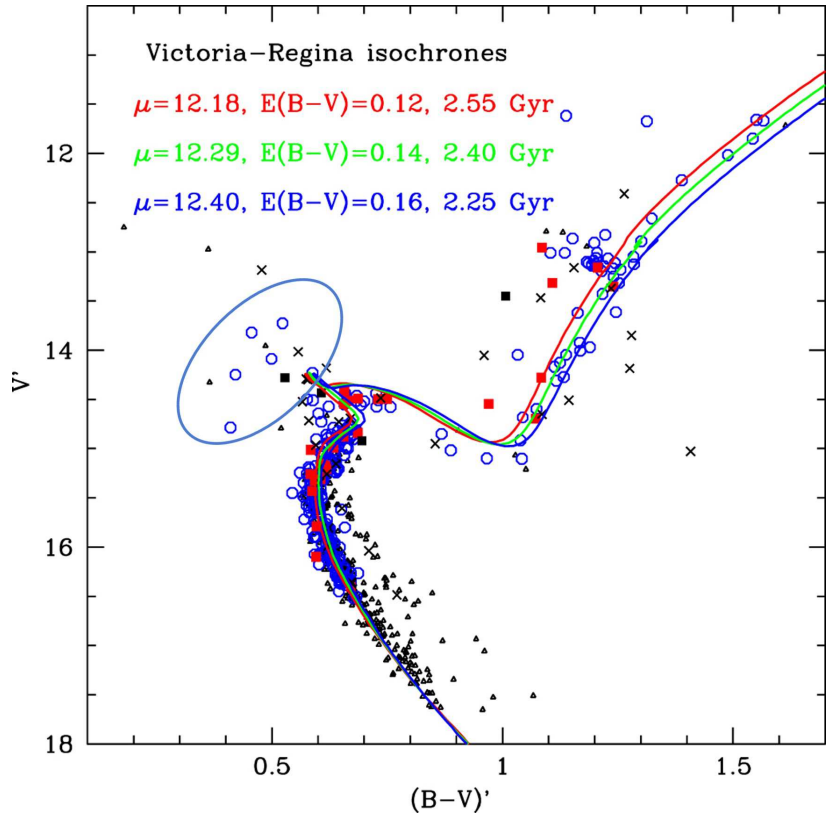}
\caption{Color-magnitude diagram of NGC 6819 from \cite{2019AJ....158..163D} with overlayed isochrones.  \copyright AAS. Reproduced with permission. Magnitudes and colors have been corrected for interstellar reddening. The blue oval, not in the original figure, encircles stars in the blue straggler region, but these are not necessarily the same stars discussed in our paper. The stars with black symbols are likely field stars and not cluster members. $\mu$ is the apparent distance modulus.}\label{fig:Deliyannis}
\end{figure}

Since the cluster is younger than the Sun, the stars at the cluster main-sequence turnoff are somewhat more massive than the Sun, near the expected mass range for $\gamma$ Doradus-type pulsating variables, which pulsate in high-order gravity modes with periods of around 1 day \citep[frequencies 0.3-3 c/d;][]{2010aste.book.....A, 2020MNRAS.491.3586L}. This cluster also contains ``blue straggler'' stars, i.e., stars on the main sequence above the cluster turnoff that should have already left the main sequence to become red giants \citep[see Fig. \ref{fig:Deliyannis} from][]{2019AJ....158..163D}. Blue stragglers are believed to have formed either via stellar mergers or mass transfer from a companion sometime in the star's past \citep{2021A&A...650A..67R}. The NGC 6819 blue stragglers have the right temperatures to show $\delta$ Scuti-type pulsations, i.e., low-order acoustic mode ($p$-mode) or gravity-mode ($g$-mode) pulsations with periods of around 2 hours \citep[frequencies 5-50 c/d;][]{2010aste.book.....A, 2015MNRAS.452.3073B}. If pulsations are found, stellar modeling and asteroseismic analysis may help to better understand the origins of these blue stragglers.  

We discuss light curves derived from {\it Kepler} NGC 6819 superstamp data and pulsation frequency analyses for five confirmed cluster members. Four stars are blue stragglers, and one is near the cluster turnoff. 

\section{{\it Kepler} Data Analysis and Results} \label{sec:Kepler}

The superstamp field centered on NGC 6819 was viewed nearly continuously for the four years (17 quarters) of the original {\it Kepler} mission.  These data span barycentric Julian Days 131.5 to 1591.0 after Julian Date 2454833.0. Gaps in the data of around 90 days for Quarters 6, 10, and 14 arise because {\it Kepler} CCD module \#3 and later \#7 (out of the total of 21) failed during the mission.

We used simple aperture photometry (SAP) pixel data for the light curves and prepared final light curves using our custom scripts and PyKE software \citep{2012PASP..124..963K}. See also \cite{2022MNRAS.509..763S} for details of choosing the apertures and optimizing them for analysis. We searched for variations in each superstamp pixel showing variability in the right range to be $\delta$ Scuti or $\gamma$ Doradus variable star candidates. This search resulted in five cluster members and eight non-members (not discussed here) for follow-up.

Cluster membership probabilities were derived using astrometry data from Gaia Data Release 3 \citep{2021A&A...649A...1G}. Table \ref{tab:astrometry} summarizes the parallax, distance, and cluster membership probability for each star. We use a 5-D approach, i.e., accounting for proper motion (2x), coordinates (2x), and parallax. See also \cite{SanjayanActaAstron} for additional details on the method to determine membership probabilities. Radial velocities are not available for our targets from Gaia DR3.  Radial velocities are available in the literature that could be used, but because they are not taken with the same instrument, instrumental shifts could result, so it was judged more reliable not to use them.

\begin{deluxetable*}{lccc}
\tablenum{1}
\tablecaption{Summary of astrometric analysis for cluster membership and distances using corrected parallaxes. The parallax fractional errors of all targets are less than 10\%.\label{tab:astrometry}}
\tablewidth{0pt}
\tablehead{
\colhead{KIC} & \colhead{Parallax} & \colhead{Distance} & \colhead{Membership}\\
\colhead{} & \colhead{[mas]} & \colhead{[pc]} & \colhead{Probability \%}
}
\startdata
5024468	& 0.414(12) & 2413(71)	& 99.5\\
5024084	& 0.384(15) & 2606(104)	& 99.3\\
5024455	& 0.475(38) & 2106(168)	& 98.1\\
5113357	& 0.422(16) & 2373(90)	& 98.9\\
5112843	& 0.451(24) & 2218(117)	& 99.1\\
\enddata
\end{deluxetable*}

Two of the five member stars discussed here, KIC 5024084 and KIC 5024455, have 14 quarters of {\it Kepler} 30-min data, and 1 month of 2-min cadence data that can be found in the Mikulski Archive for Space Telescopes (MAST).\footnote{MAST, https://archive.stsci.edu/} We compared amplitude spectra produced using light curves from our custom analysis and those using the Pre-search Data Conditioning Simple Aperture Photometry (PDC\_SAP) 30-min-cadence light curves from MAST.  We find that the signal-to-noise ratio (S/N) is marginally better using the PDC\_SAP data, whereas the frequencies found were the same. For thees two stars, we therefore used the PDC\_SAP long-cadence data for the amplitude spectra and frequency tables presented here.

Some of these stars also have been observed by the Transiting Exoplanet Survey Satellite ({\it TESS}) mission \citep{2015JATIS...1a4003R}. The {\it Kepler} pixel dimensions are 4 $\times$ 4 arcsec, considerably smaller than the 21 arcsec pixels of {\it TESS}, reducing the risk and consequences of contamination of the light curves by nearby stars in the crowded field in the cluster center.

To determine significant frequencies, the light curves were processed by Fourier analysis, and the successive highest-amplitude peaks removed from the light curve by pre-whitening until only noise remained.  We used a detection threshold S/N around 5 as discussed by \cite{2015MNRAS.448L..16B}. This level is the average over the entire {\it Kepler} frequency spectrum from 0 c/d to the Nyquist frequency limit of 24.4695 c/d for 30-min cadence data. Generally, the noise level is greater for low frequencies than for higher frequencies in this range.

For {\it Kepler} time series longer than about one year, frequencies higher than the Nyquist limit can be found by taking advantage of the shift in arrival time of the signal caused by the light travel time to the spacecraft in orbit around the Sun \citep{2012MNRAS.424.2686B, 2013MNRAS.430.2986M}. The true frequencies have higher amplitudes than their Nyquist-reflected frequencies in the amplitude spectrum (see examples in sub-sections 2.1 and 2.4).

Table \ref{tab:properties} summarizes properties of the five cluster member stars discussed in the remainder of this Section. The effective temperature, log surface gravity, radius, mass, and luminosity are from the {\it TESS} Input Catalog (TIC) version 8.2 \citep{2019AJ....158..138S} available on MAST. The stellar quantities are approximate as many are derived from color photometry and stellar model grids. Spectroscopy and asteroseismic modeling making use of the stellar pulsation frequencies should improve the accuracy of these quantities.

While finalizing this paper, we found that \cite{2022ApJS..258...39C} presented light curves of stars in the NGC 6819 superstamp field using Increased Resolution Image Subtraction photometry. The Ph.D. thesis of \cite{2020PhDT........26C} shows amplitude spectra derived for the five stars we discuss here and classifies them as $\delta$ Sct or $\gamma$ Dor variables. Our work has been performed independently from \cite{2022ApJS..258...39C} and contains additional analysis, including tables of significant frequencies, new spectroscopic results, and analysis of frequency or period spacings for several stars. We note in Table \ref{tab:properties} the variable star classifications of \cite{2020PhDT........26C}, which in a few cases differ from our own.

\subsection{KIC 5024468}

KIC 5024468 was known pre-{\it Kepler} as a $\delta$ Sct variable in the blue straggler region of the NGC 6819 color-magnitude diagram \citep{2010AJ....140.1268T}. The {\it Kepler} light curve for this star is contaminated by light from a nearby eclipsing binary, KIC 5024450, with a period of 3.05 days. 

\begin{figure*}
\gridline{\fig{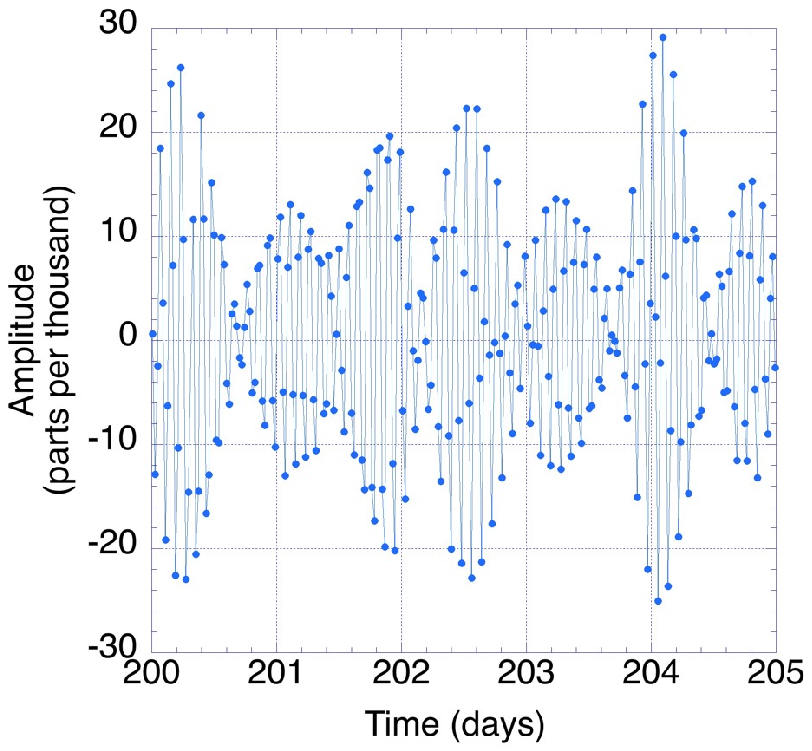}{0.3\textwidth}{(a)}
          \fig{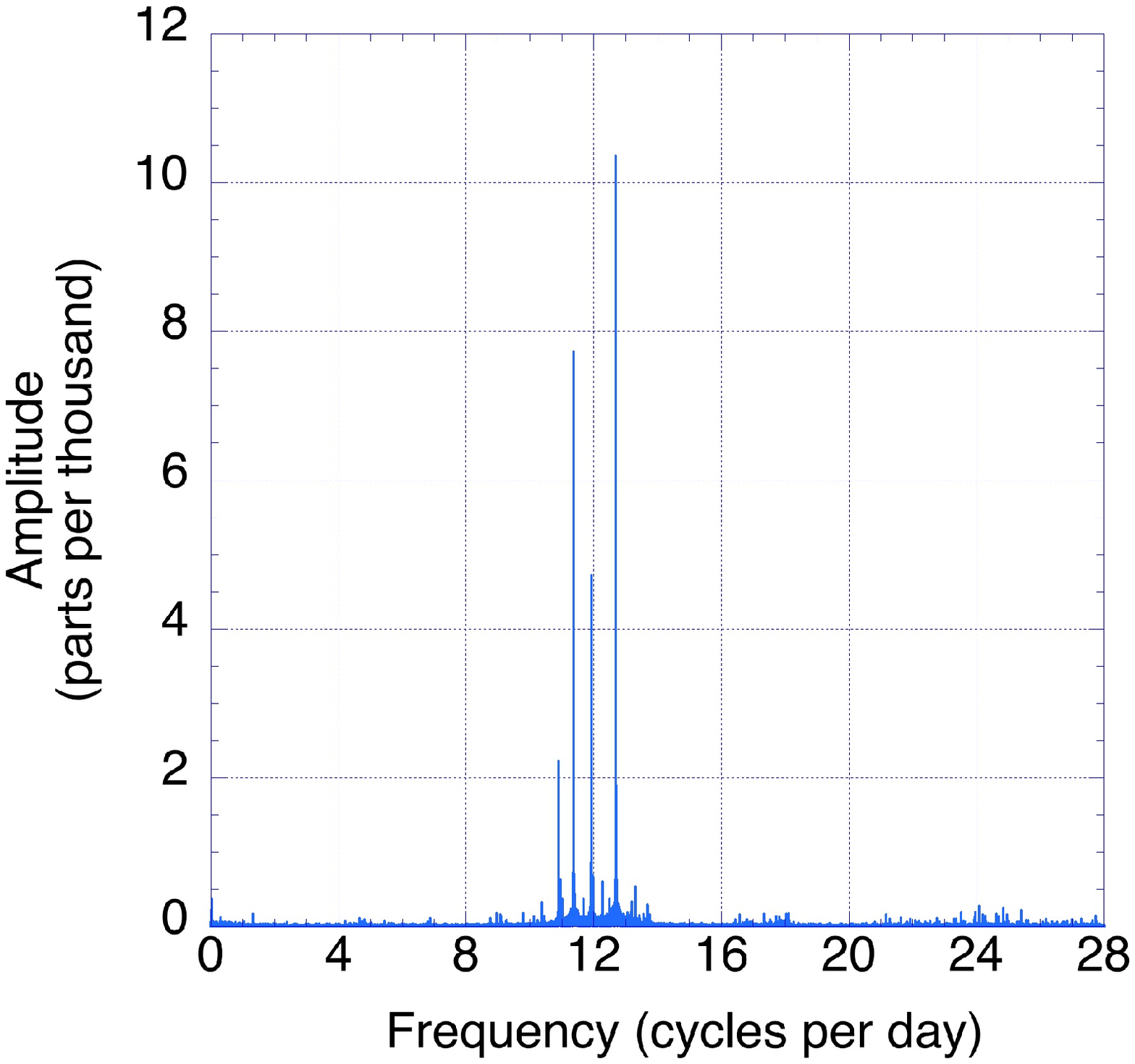}{0.3\textwidth}{(b)}
          \fig{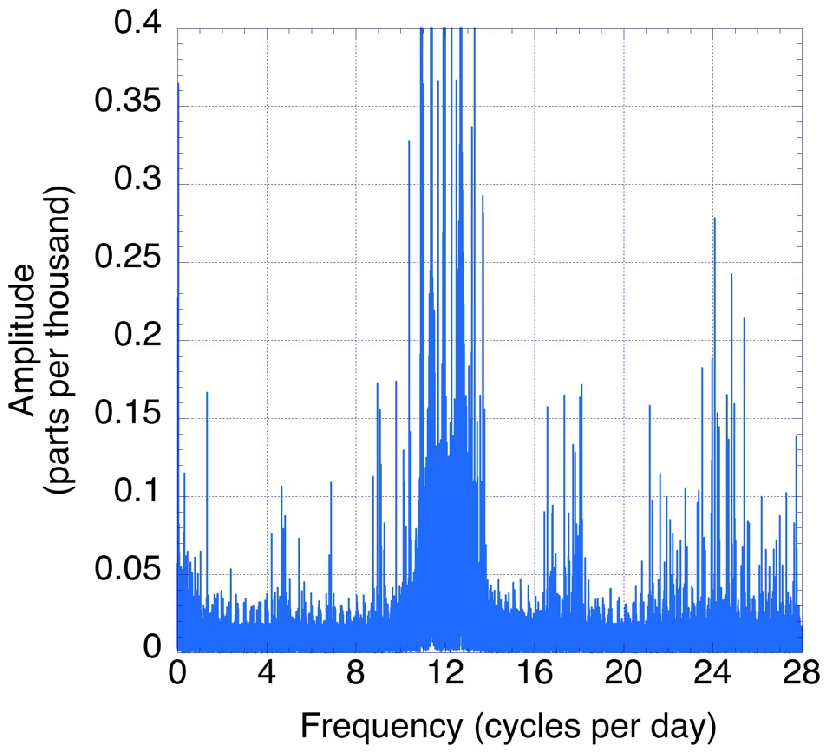}{0.3\textwidth}{(c)}
          }
\caption{(a) Zoom-in on 5-day portion of KIC 5024468 light curve. (b) Amplitude spectrum for KIC 5024468 showing $\delta$ Sct modes.  The Nyquist frequency for 30-min {\it Kepler} cadence data is 24.4695 c/d, but the spectrum extends to 28 cycles/day as there are real super-Nyquist frequencies that have higher amplitudes than their reflections. (c) Zoom-in on low-amplitude portion of KIC 5024468 spectrum showing many significant low-amplitude modes in both the $\gamma$ Dor and $\delta$ Sct frequency ranges.}
\label{fig:KIC5024468}
\end{figure*}

Figure \ref{fig:KIC5024468} (left) shows a 5-day zoom-in on the KIC 5024468 light curve; Figure \ref{fig:KIC5024468} (center) shows the amplitude spectrum, revealing many modes in the $\delta$ Sct frequency range; the highest-amplitude modes have frequencies around 12 c/d. Pre-whitening analysis reveals 236 significant frequencies with S/N $>$ 6, including 7 real frequencies above the Nyquist limit (Table \ref{tab:A1}). Figure \ref{fig:KIC5024468} (right) zooms in on the low-amplitude portion of the spectrum, showing several frequencies in the $\gamma$ Dor frequency range, with frequencies $<$ 5 c/d. We therefore categorize this star as a $\delta$ Sct/$\gamma$ Dor hybrid candidate.

\begin{figure*}
\gridline{\fig{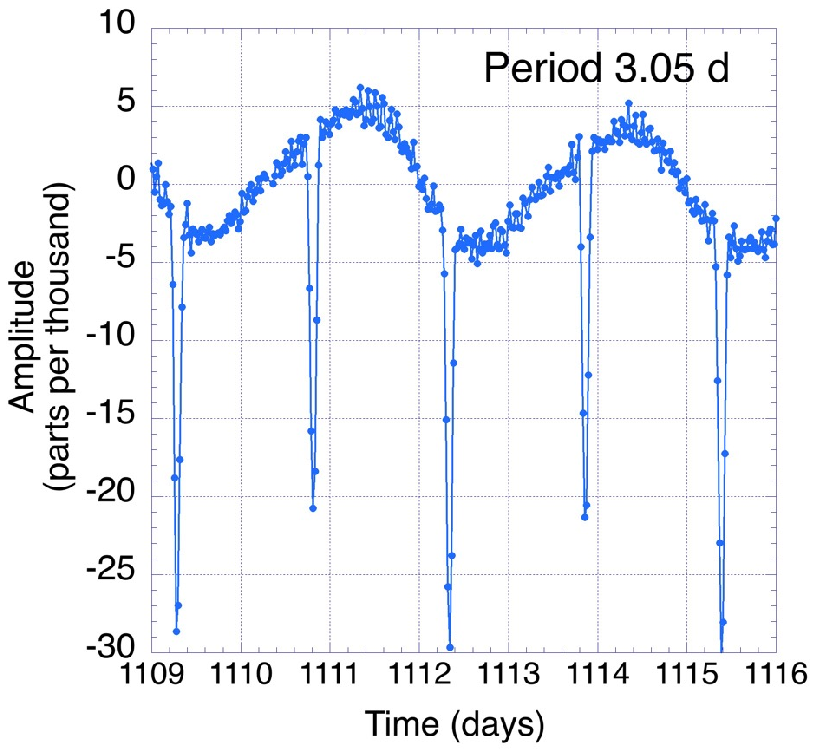}{0.4\textwidth}{(a)}
          \fig{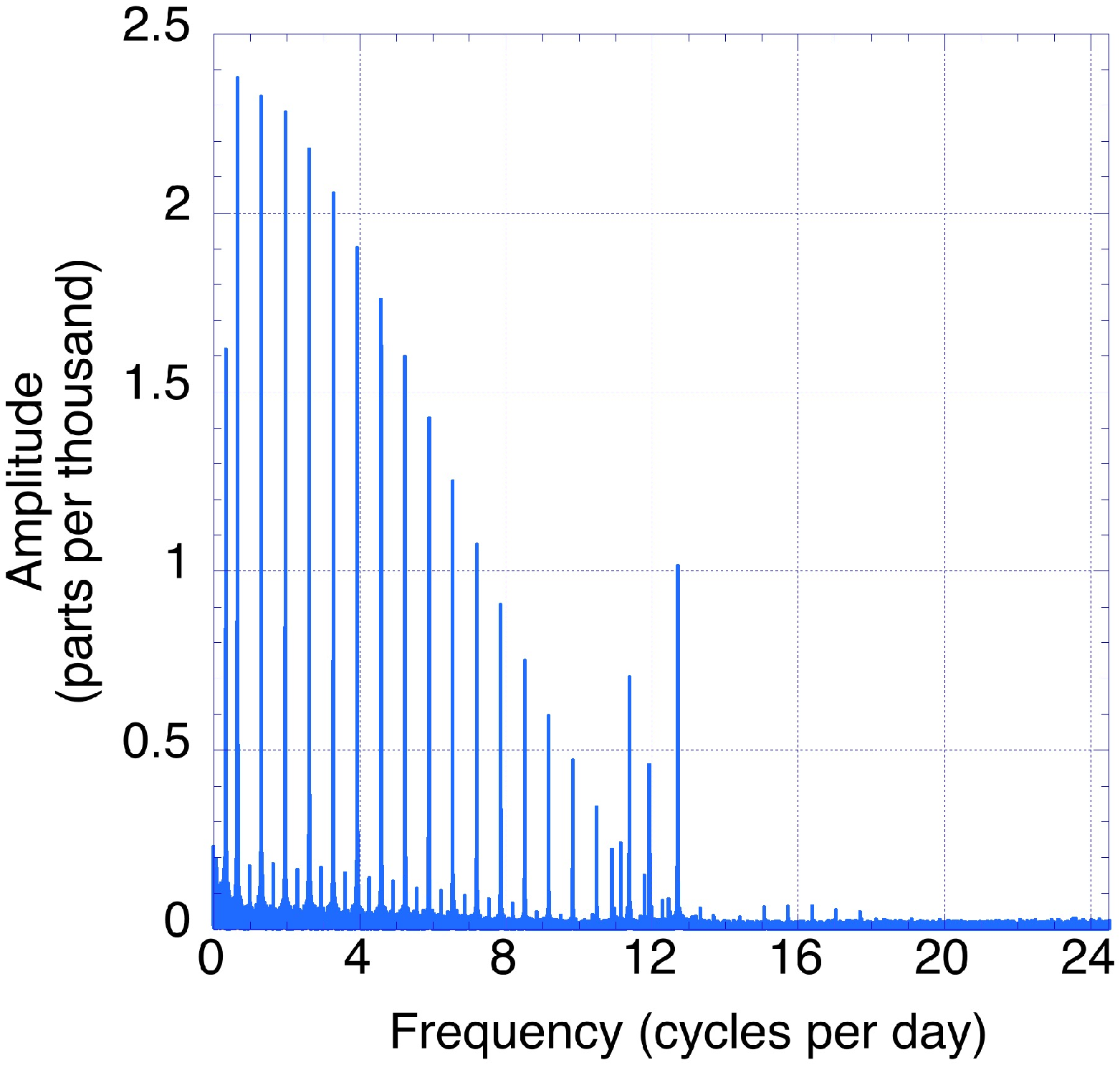}{0.4\textwidth}{(b)}
          }
\caption{(a) Zoom-in on 5-day portion of light curve of eclipsing binary KIC 5024450, showing eclipses with period 3.05 d. This light curve is contaminating the light curve of 5024468, and the binary orbital frequency and its harmonics were removed when determining the frequencies of KIC 5024468 in Table \ref{tab:A1}. (b) Amplitude spectrum of KIC 5024450, showing characteristic comb-like structure for Fourier analysis of an eclipsing binary. The $\delta$ Sct oscillations of KIC 5024468 around 12 c/d also contaminate the KIC 5024450 spectrum.}
\label{fig:KIC5024450}
\end{figure*}

Figure \ref{fig:KIC5024450} shows the light curve and amplitude spectrum of the eclipsing binary KIC 5024450, which is contaminating the light curve of KIC 5024468.  As can be seen in Fig.\,\ref{fig:KIC5024450}(b), the $\delta$ Sct oscillations of KIC 5024468 around 12 c/d contaminate the KIC 5024450 spectrum; the binary frequency and its harmonics were removed when determining the KIC 5024468 frequencies in Table \ref{tab:A1}.

\subsection{KIC 5024084}

KIC 5024084 is listed as a blue straggler in the SIMBAD\footnote{http://simbad.u-strasbg.fr/simbad/} database. The {\it Kepler} light curve available via MAST for this star was studied using long-cadence data through Quarter 12 and one month of short-cadence data by \cite{2013MNRAS.430.3472B}. According to their description, ``KIC 5024084 shows very clear variations with irregular amplitudes but a distinct period of 2.07 d which is probably the rotation period of the star with spots.''

\begin{figure*}
\gridline{\fig{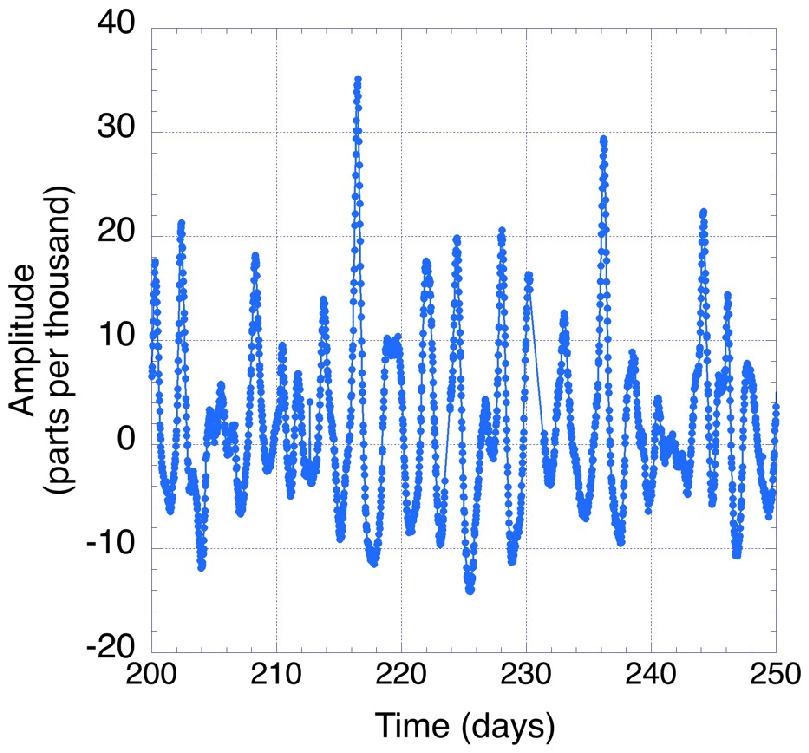}{0.3\textwidth}{(a)}
          \fig{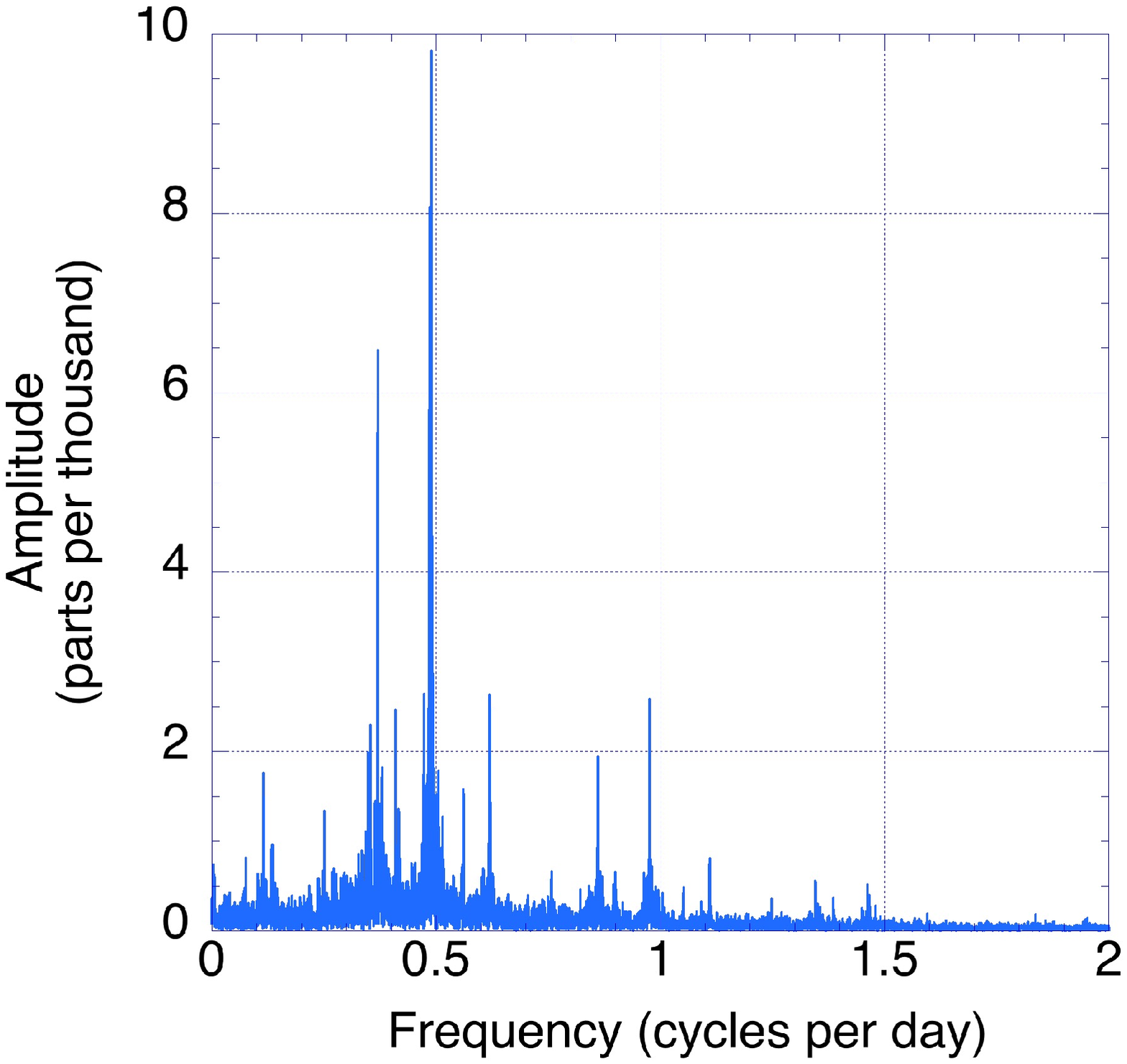}{0.3\textwidth}{(b)}
          \fig{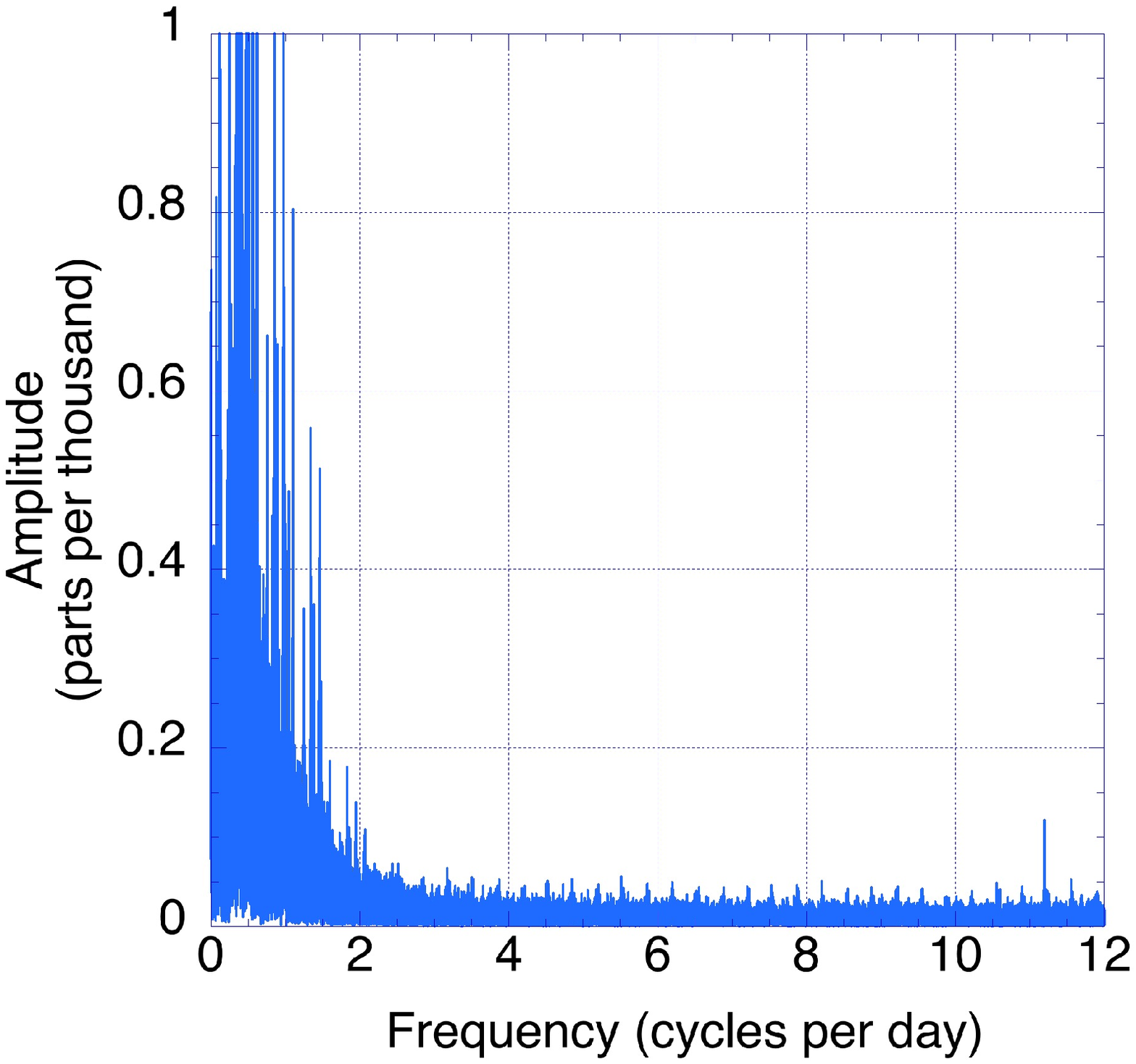}{0.3\textwidth}{(c)}
          }
\caption{(a) Zoom-in on 50-day portion of KIC 5024084 light curve. (b) Low-frequency portion of KIC 5024084 amplitude spectrum, showing many low-frequency modes. (c) Zoom-in on low amplitudes of KIC 5024084 amplitude spectrum, with frequency range extended to 12 c/d.  A single 11.2 c/d mode is visible in $\delta$ Sct frequency range.}
\label{fig:KIC5024084}
\end{figure*}

Figure \ref{fig:KIC5024084} (left) shows a 50-day zoom in of the KIC 5024084 light curve extracted from the superstamp pixels. Figure \ref{fig:KIC5024084} (center) shows the amplitude spectrum, revealing many low-frequency modes, including the two highest-amplitude modes with periods 2.038 and 2.058 days that may correspond to the period identified as a probable rotation period by \cite{2013MNRAS.430.3472B}. \cite{2015AA...583A..65R} list KIC 5024084 in their catalog of 12,319 {\it Kepler} stars with multiple peaks, which they interpret as differential rotation.  Their catalog lists 2.038 c/d and 2.060 c/d as the minimum and maximum period range, coinciding with the two highest-amplitude peaks that we find.

It is not straightforward to determine the origin of low-frequency peaks in the amplitude spectrum.  \cite{2018MNRAS.474.2774S} offer an alternative explanation for `hump and spike' features in amplitude spectra seen in some $\gamma$ Dor stars as a rotation frequency (spike) accompanied by a lower-frequency cluster of global Rossby modes (hump). Other groupings of modes, e.g., some higher than the rotation frequency, may be $\gamma$ Dor gravity modes. For KIC 5024084, there may also be harmonics of the largest-amplitude modes (around 0.5 c/d) at around 1 c/d and 1.5 c/d.  Pre-whitening analysis revealed 53 modes (Table \ref{tab:A2}).  Figure \ref{fig:KIC5024084} (panel c) shows the low-amplitude portion of the spectrum extended to 12 c/d; the mode visible in the $\delta$ Sct frequency range at 11.2 c/d has S/N = 25.  Three additional $\delta$ Sct modes with lower S/N ratio are revealed in the pre-whitening analysis and listed in Table \ref{tab:A2}.  We categorize this star as a $\gamma$ Dor/$\delta$ Sct hybrid candidate.

A very close inspection of Figure \ref{fig:KIC5024084} (panel c) shows a comb of low-amplitude peaks with frequency spacing around 1/3 c/d.  There is a known artifact in the {\it Kepler} data at this frequency from thruster firings every 3.0 days to desaturate angular momentum buildup in the reaction wheels (see {\it Kepler} Data Release 3 Notes, KSCI-19043-001).\footnote{https://ntrs.nasa.gov/api/citations/20100027540/downloads/20100027540.pdf}  For this star, we analyzed the PDC\_SAP light curve from MAST, and it is possible that this artifact was not completely cleaned from the data.  This low-amplitude comb is no longer visible in the residual after pre-whitening the highest amplitude frequency.


\subsection{KIC 5024455}

KIC 5024455 is also listed as a blue straggler in SIMBAD. This star's {\it Kepler} light curve data were studied using early data releases by \cite{2011A&A...534A.125U}, who categorize it as a $\gamma$ Dor star, and later by \cite{2013MNRAS.430.3472B} using long-cadence data up through Quarter 12 and 1 month of short-cadence data, who categorize it as a suspected $\gamma$ Dor star. \cite{2014AJ....148...38M} list KIC 502445 (also known as WOCS 014012) as a single-line spectroscopic binary with a 762-day orbital period.

\begin{figure*}
\gridline{\fig{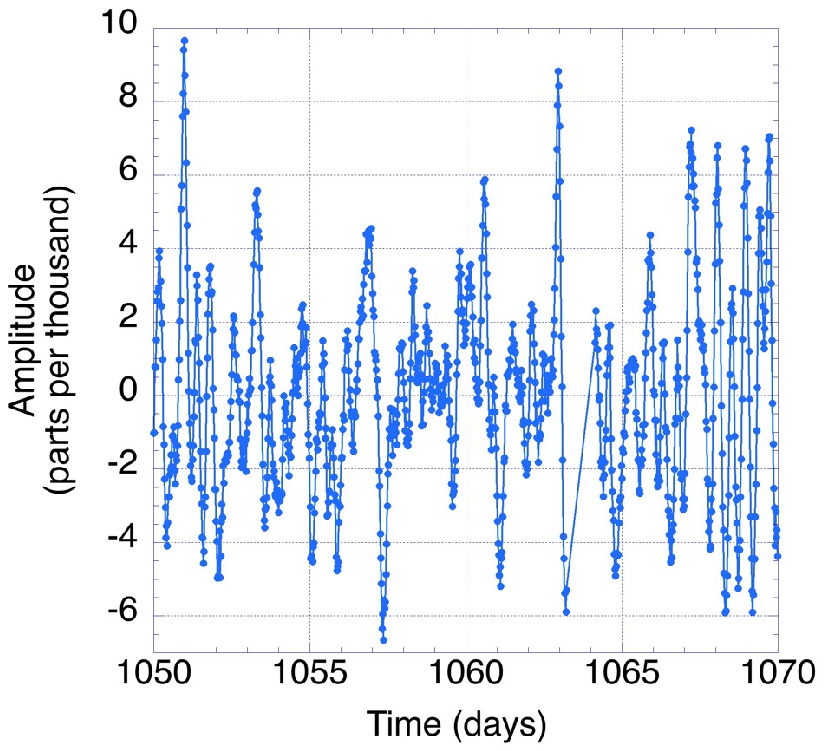}{0.4\textwidth}{(a)}
          \fig{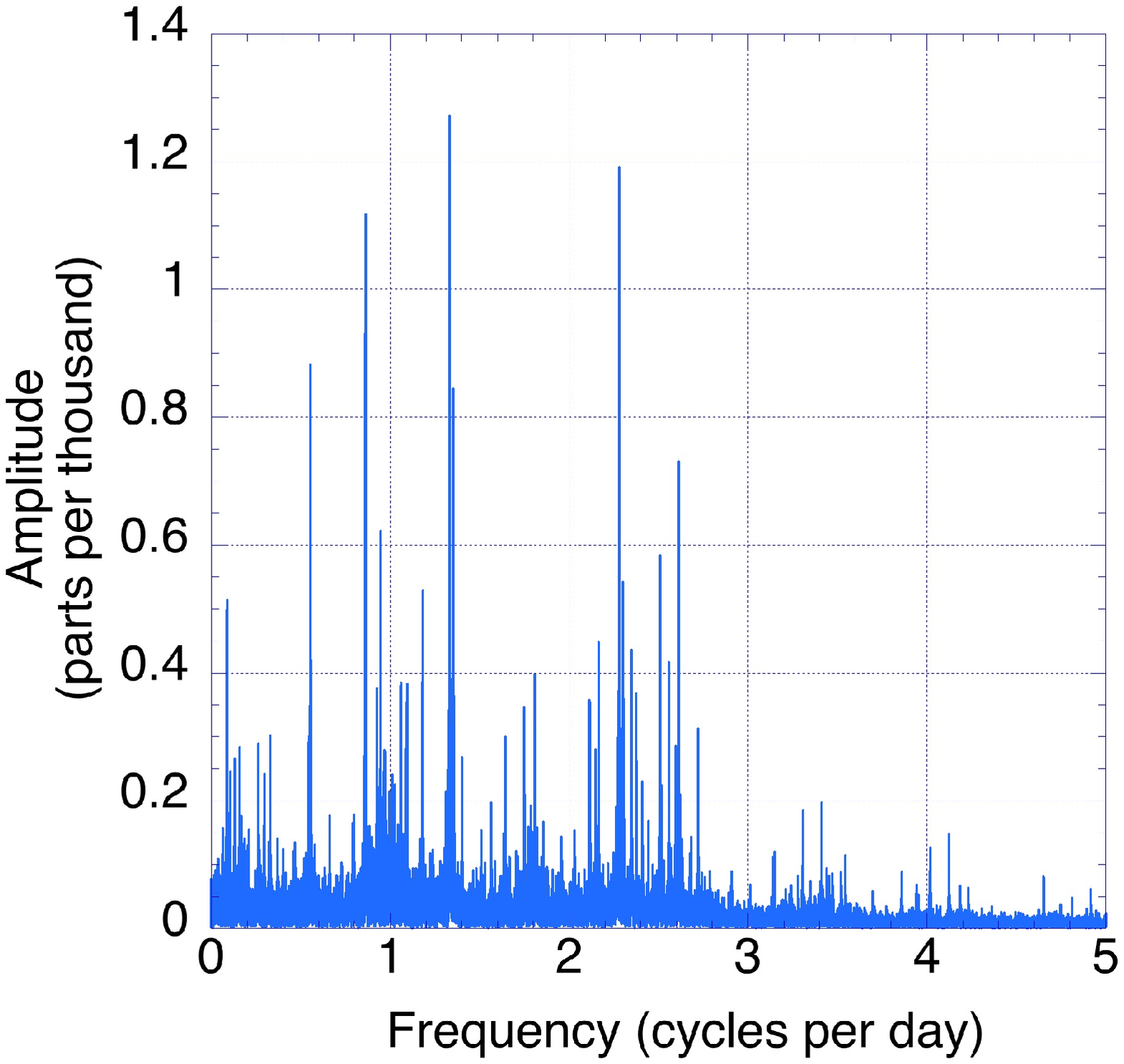}{0.4\textwidth}{(b)}
          }
\caption{(a) Zoom-in on 20-day portion of KIC 5024455 light curve. (b) KIC 5024455 amplitude spectrum.}
\label{fig:KIC5024455}
\end{figure*}

Figure \ref{fig:KIC5024455} (left) shows a 20-day zoom-in on a portion of the light curve. The amplitude spectrum (Fig. \ref{fig:KIC5024455}, right) shows only modes with frequencies $<$ 5 c/d, in the $\gamma$ Dor frequency range. Pre-whitening analysis reveals 84 frequencies with S/N $>$ 8.4 (Table \ref{tab:A3}).  There may be additional significant frequencies, but it is difficult to be sure they are real because of higher noise levels at low frequencies. While some of these low-frequency groupings may be gravity modes, some may also be global Rossby modes, and more analysis will be needed in conjunction with stellar models to understand and identify the mode patterns. Pulsation modes cannot be distinguished from signatures of rotation and star spots using the light curve alone. We categorize this star as a $\gamma$ Dor candidate.

\vspace{15pt}

\subsection{KIC 5113357}

\begin{figure*}
\gridline{\fig{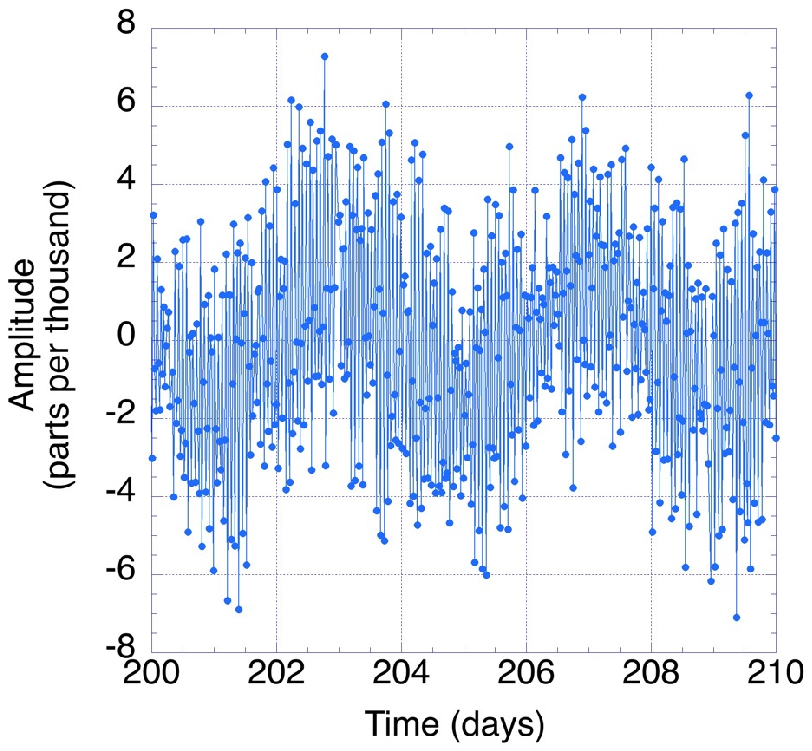}{0.4\textwidth}{(a)}
          \fig{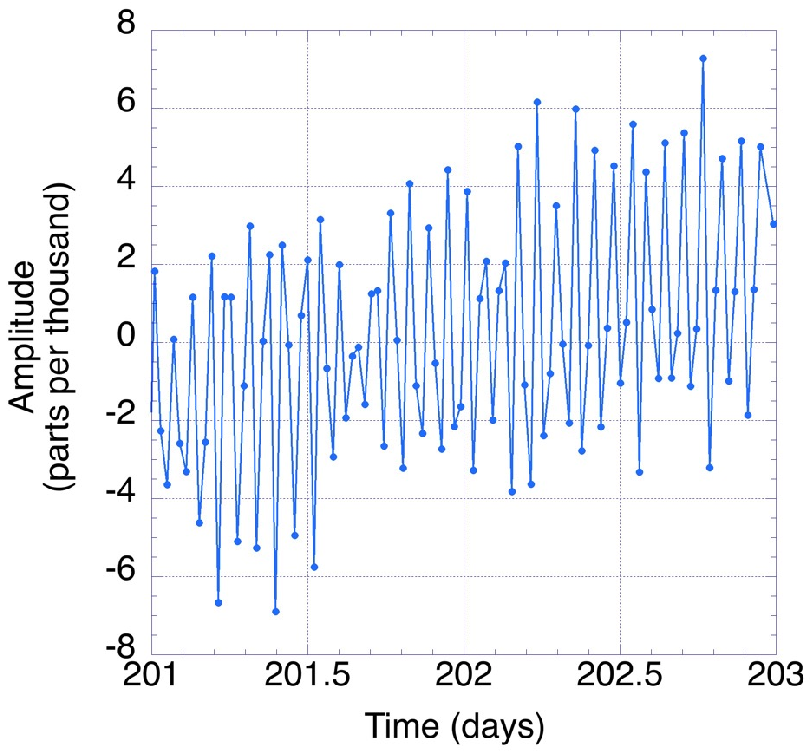}{0.4\textwidth}{(b)}
          }
\caption{(a) Zoom-in on 10-day portion of KIC 5113357 light curve. (b) Zoom-in on 2-day portion of KIC 5113357 light curve.}
\label{fig:KIC5113357lc}
\end{figure*}

\begin{figure*}
\gridline{\fig{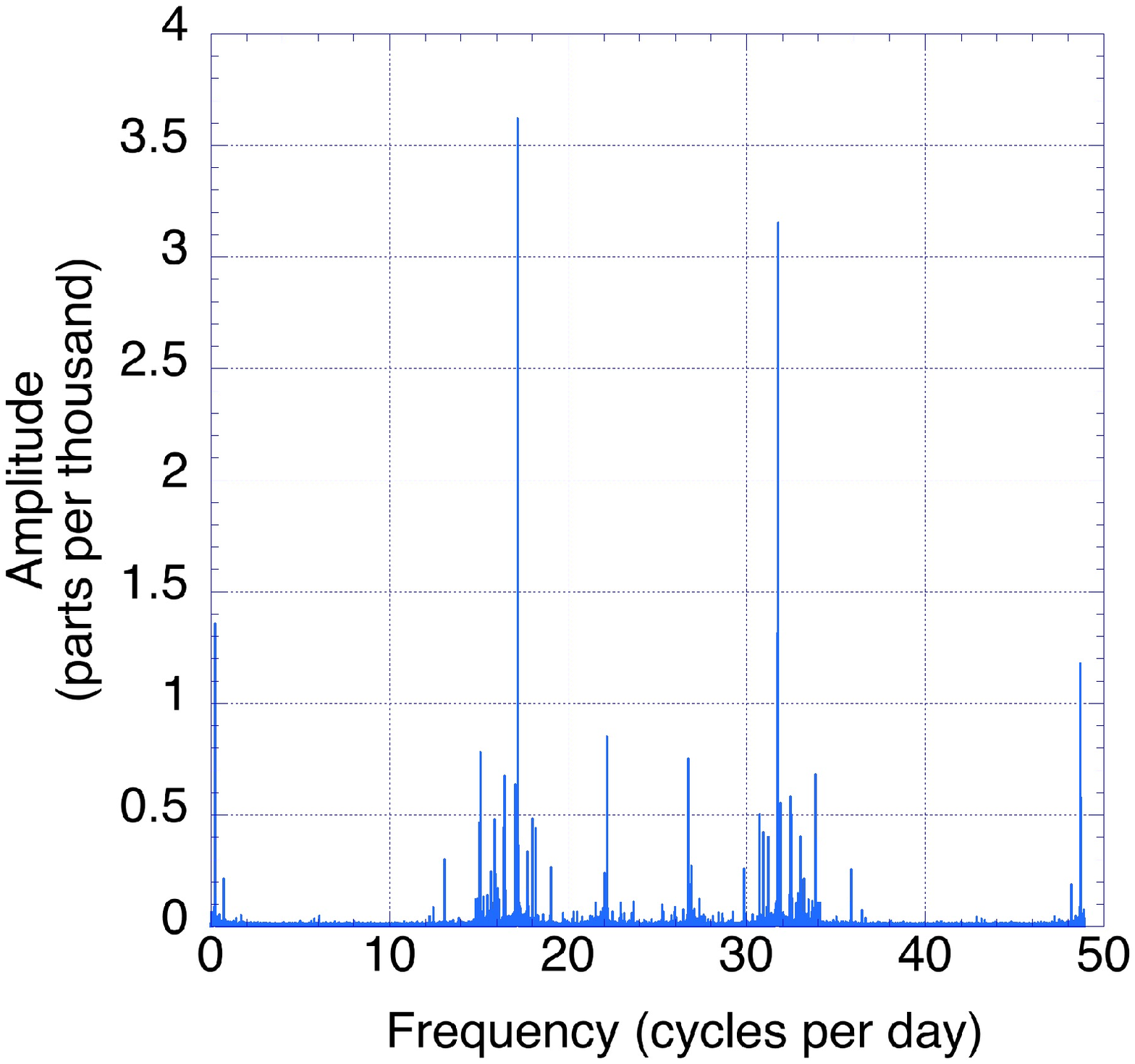}{0.4\textwidth}{(a)}
          \fig{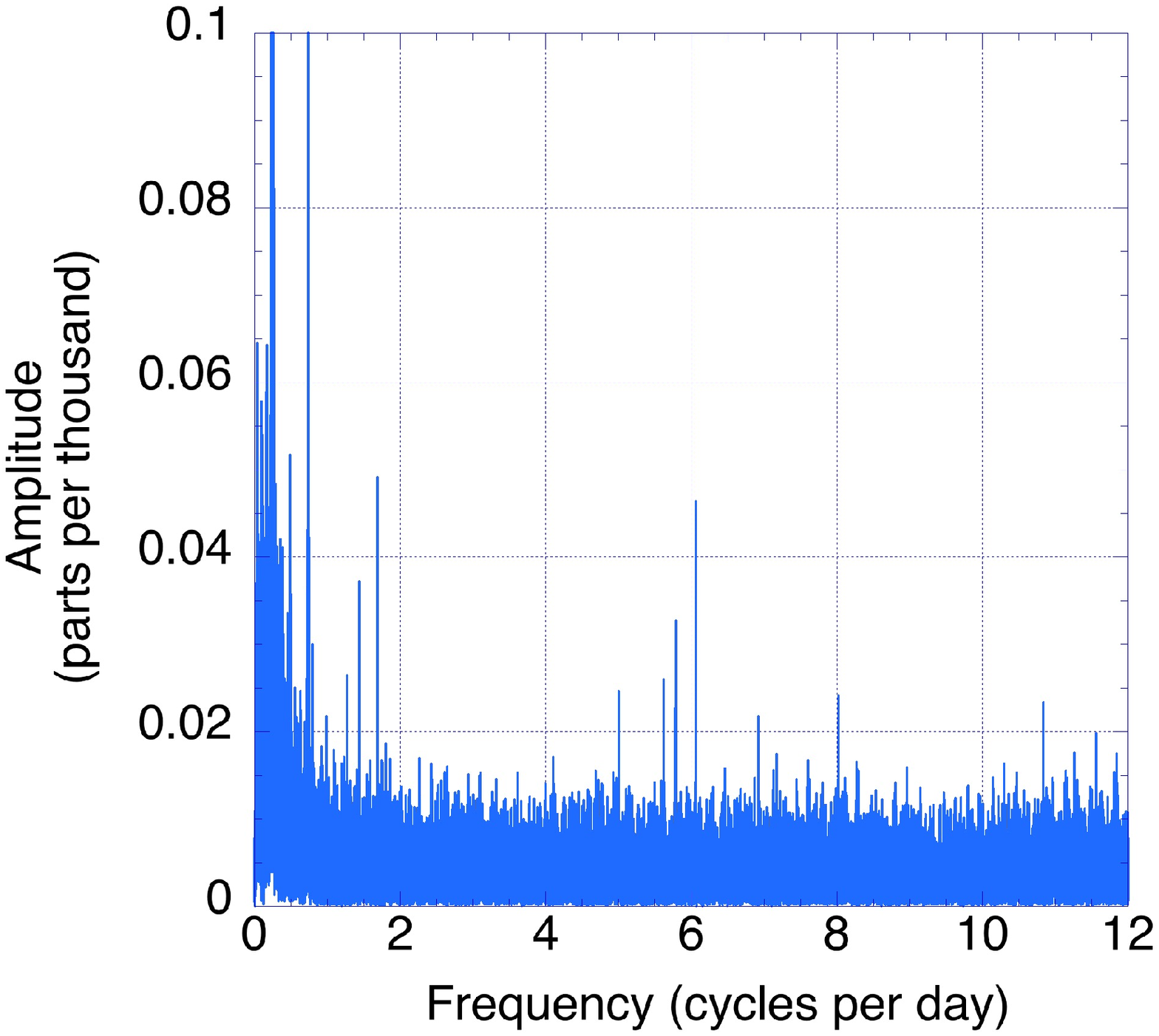}{0.4\textwidth}{(b)}
          }
\caption{(a) KIC 5113357 amplitude spectrum, extended to 50 c/d. Some frequencies above the Nyquist frequency of 24.4695 c/d have higher amplitudes than their reflections and are real frequencies. (b) Zoom-in on low-amplitude and low-frequency portion of KIC 5113357 amplitude spectrum.  A few low-amplitude frequencies are present in the $\gamma$ Dor frequency range.}
\label{fig:KIC5113357ft}
\end{figure*}

KIC 5113357 does not have {\it Kepler} data available in MAST, and it is not categorized as a variable star in SIMBAD.  Its temperature and luminosity do place it in the blue straggler region for the NGC 6819 cluster.

Figure \ref{fig:KIC5113357lc} (left) shows a 10-day zoom-in on the KIC 5113357 light curve derived from the superstamp pixel data. There is an overall modulation at 0.24755 c/d (period around 4 days); this low frequency is the 2nd highest peak in the amplitude spectrum (see Fig. \ref{fig:KIC5113357ft}). This peak could be a binary orbital frequency, and this star could be a contact eclipsing binary.  This frequency could also be a rotational frequency. Table \ref{tab:A4} lists this frequency first, followed by three harmonics that are also found in the amplitude spectrum.

Figure \ref{fig:KIC5113357lc} (right) shows a 2-day zoom-in on the light curve, revealing higher frequency oscillations in the $\delta$ Sct frequency range. Figure \ref{fig:KIC5113357ft} (left) shows the amplitude spectrum out to 50 c/d.  It is evident that most frequencies are reflected about the Nyquist frequency limit of 24.4695 c/d.  However, some super-Nyquist frequencies have larger amplitudes than their reflected counterparts, and are the real frequencies. Apart from the 0.24755 c/d period mentioned above, pre-whitening of the spectrum shows 120 additional modes with S/N $>$ 5.3, most in the $\delta$ Sct frequency range (Table \ref{tab:A4}). 36 of the 120 frequencies are above the Nyquist limit, with the highest of these, 36.496 c/d, still in the $\delta$ Sct range. Figure \ref{fig:KIC5113357ft} (right) shows a zoom-in on the low-amplitude and low-frequency portion of the amplitude spectrum. There are a few modes in the $\gamma$ Dor frequency range. We therefore categorize this star as a $\delta$ Sct/$\gamma$ Dor hybrid candidate.

\subsection{KIC 5112843}

\begin{figure*}
\gridline{\fig{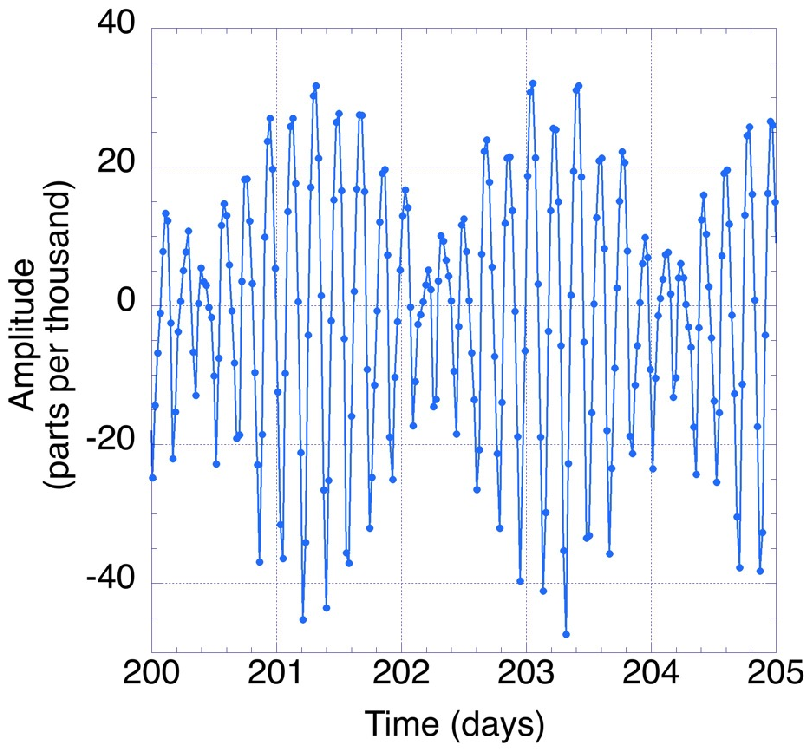}{0.4\textwidth}{(a)}
          \fig{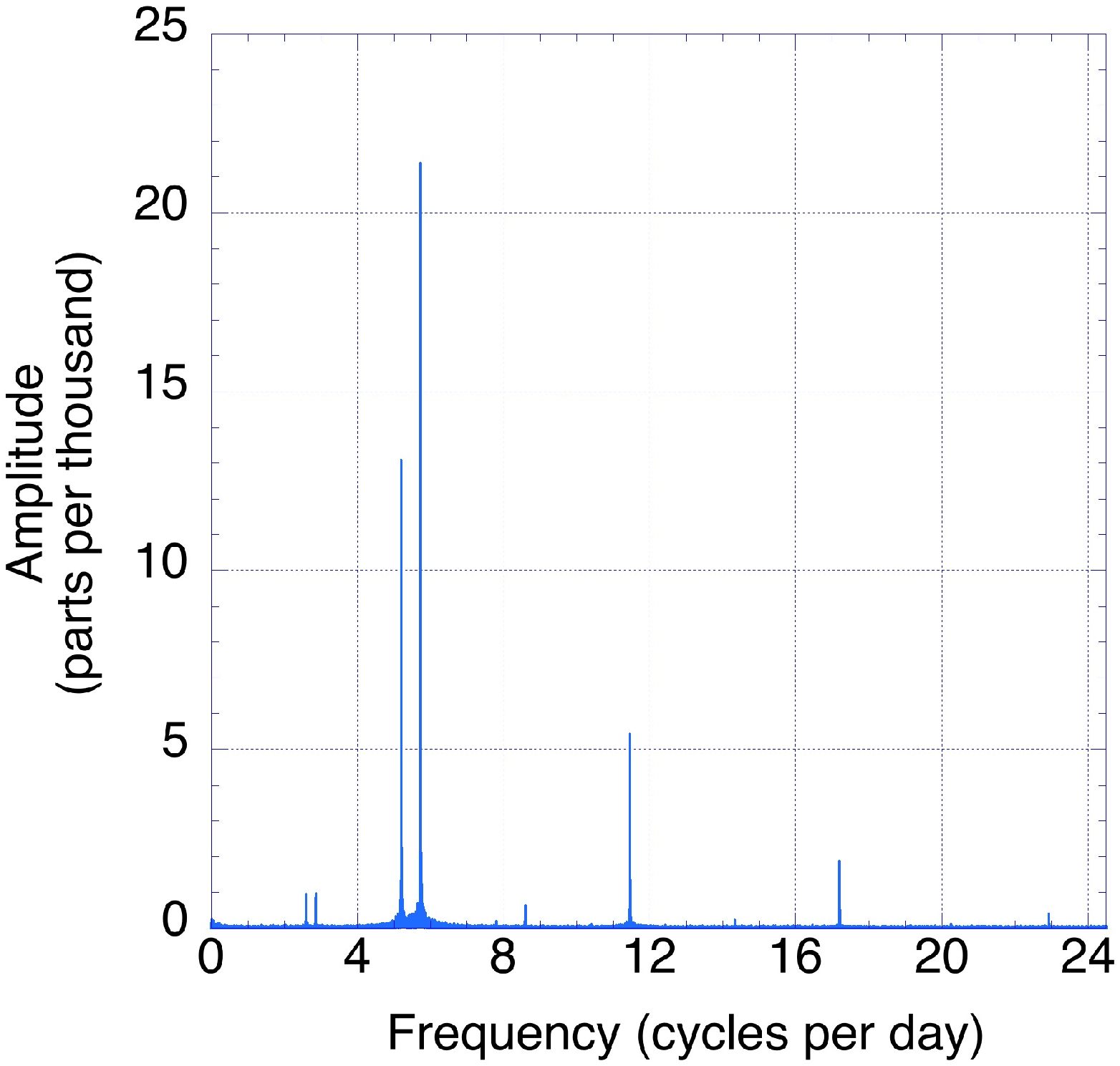}{0.4\textwidth}{(b)}
          }
\caption{(a) 5-day zoom-in on KIC 5112843 light curve. The dominant property is two close frequencies beating against other. (b) KIC 5112843 amplitude spectrum.}
\label{fig:KIC5112843}
\end{figure*}

\begin{figure}
\centering
\includegraphics[width=0.5\textwidth]{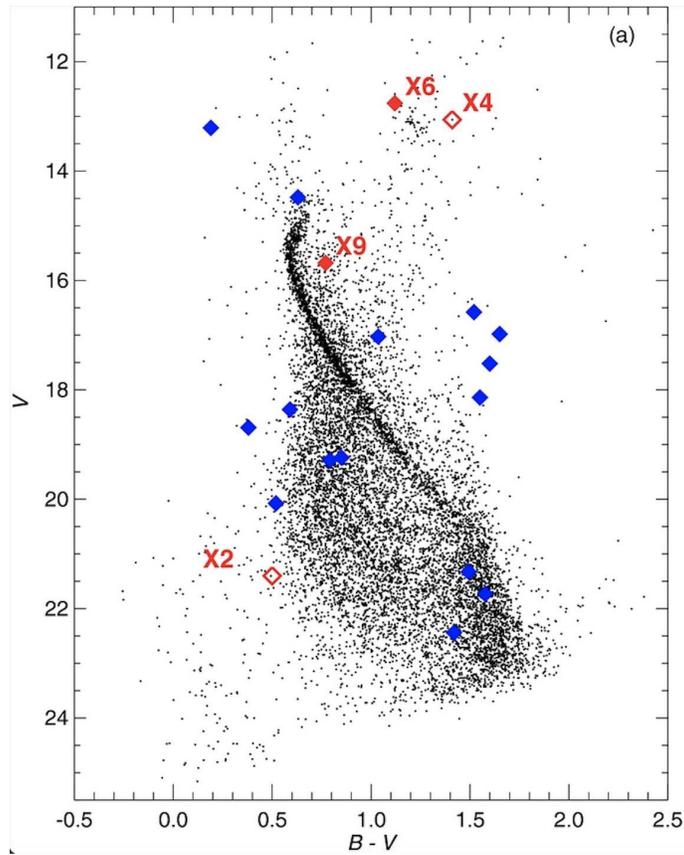}
\caption{Figure 5a from \cite{2012ApJ...745...57G}, which shows NGC 6819 x-ray binaries (red symbols) in the color-magnitude diagram. X9 is KIC 5112843, located near the main-sequence turnoff.  \copyright AAS. Reproduced with permission. \label{fig:Gosnell}}
\end{figure}

The last of the five NGC 6819 members we discuss is another interesting and mysterious star, KIC 5112843. This star does not have processed {\it Kepler} data available in MAST. It can be found in SIMBAD under its TIC catalog number, and it is listed as an eclipsing binary. \cite{2010AJ....140.1268T} discuss the ground-based light curve of this star from pre-{\it Kepler} data, and they note that sometimes the light curve shows one shallow, almost nonexistent, dip that causes it to resemble a detached eclipsing binary. At other times, the light curve resembles that of a contact binary. Concerning these latter phases, \cite{2010AJ....140.1268T} write, ``the system showed some of its deepest eclipses and showed variations identifying it as an EW [also known as W UMa] contact binary system, with gravitationally distorted stars of nearly equal temperature.''

The {\it Kepler} light curve derived from the superstamp pixel data shows the signal of two close frequencies beating against each other, combining constructively or destructively to create the pattern seen in Figure \ref{fig:KIC5112843}. The amplitude spectrum shows the two largest-amplitude modes that are close in frequency at 5.2071 and 5.7357 c/d, but also two lower-amplitude frequencies at half these values, 2.6036 and 2.8678 c/d. Pre-whitening analysis of the light curve reveals only harmonics of these two frequencies, six of them of the 2.6036 c/d mode, and nine of the 2.8678 c/d mode (Table \ref{tab:A5}). \cite{2010AJ....140.1268T} identify the binary period as 0.348687 d (frequency 2.8679 c/d), corresponding to one of the lower-amplitude frequencies in the amplitude spectrum. It is not known why there are pairs of frequencies, and why the amplitudes of the parent frequencies are smaller than their first harmonic.

KIC 5112843 also was the target of x-ray observations by the XMM-Newton space telescope \citep{2012ApJ...745...57G}. \cite{2012ApJ...745...57G} find that this star is an x-ray source, and list it as an active binary. Figure \ref{fig:Gosnell} from \cite{2012ApJ...745...57G} shows the location of NGC 6819 x-ray sources on the color-magnitude diagram, including KIC 5112843 labeled as X9, superimposed on photometry from \cite{2001AJ....122..266K}. \cite{2012ApJ...745...57G} speculate that this star is a possible sub-subgiant binary system, similar to RS CVn binary systems, which are defined as close but detached binaries with active chromospheres that can cause large star spots \citep{1979ApJ...227..907E}.

\begin{deluxetable*}{lccccc}
\tablenum{2}
\tablecaption{Summary of properties of five NGC 6819 stars observed in {\it Kepler} superstamp field. Effective temperature, log surface gravity, radius, mass, luminosity, and distance are from the {\it TESS} Input Catalog (TIC) version 8.2 (Stassun et al. 2019) available on MAST.\label{tab:properties}}
\tablewidth{0pt}
\tablehead{
\colhead{ } & \colhead{KIC 5024468\tablenotemark{a}} & \colhead{KIC 5024084\tablenotemark{b}} & \colhead{KIC 5024455\tablenotemark{c}}& \colhead{KIC 5113357\tablenotemark{d}} & \colhead{KIC 5112843\tablenotemark{e}}
}
\startdata
RA (deg)&295.3227668&295.2647409&295.3210065&295.4430993&295.3576433 \\
DEC (deg)&40.18431117&40.14515408&40.10113114&40.27555384&40.20623482 \\
TIC ID&1880383370&139109448&139109202&184010448&139154029 \\
V (mag)&12.983 ± 0.046&14.874 ± 0.15&14.943 ± 0.046&14.971 ± 0.183&15.772 ± 0.126 \\
$T\rm_{eff}$ (K)&7059 ± 130&6501 ± 123&6701 ± 122&7328 ± 122&5493 ± 126 \\
\loggcms&3.442 ± 0.096&3.802&4.246&4.142&3.777 \\
Radius (R$_{\odot}$)&3.93 ± 0.26&2.40&1.49&1.81&2.10 \\
Mass (M$_{\odot}$)&1.56 ± 0.25&1.33&1.42&1.66&0.96 \\
Luminosity (L$_{\odot}$)&34.57 ± 4.03&9.267 & 4.014 &8.519 &3.607  \\
\hline
HRD location&Blue straggler &Blue straggler&Blue straggler&Blue straggler&Near main-sequence turnoff \\
Number of Frequencies &236& 53 &84&124&17 \\
\hline
Classification&$\delta$ Sct/$\gamma$ Dor &$\gamma$ Dor/$\delta$ Sct 
&$\gamma$ Dor &$\delta$ Sct/$\gamma$ Dor &Eclipsing  \\
&hybrid candidate&hybrid candidate&candidate&hybrid candidate&binary \\
\hline
\hline
\enddata
\tablenotetext{a}{Light curve contaminated by KIC 5024450; SIMBAD: $\delta$ Sct variable; \cite{2010AJ....140.1268T}: $\delta$ Sct star; \cite{2020PhDT........26C}: $\delta$ Sct star.}
\tablenotetext{b}{SIMBAD: Blue straggler; \cite{2013MNRAS.430.3472B}: Rotational variable with spots; \cite{2015AA...583A..65R}: Differentially rotating variable; \cite{2020PhDT........26C}: $\gamma$ Dor star.}
\tablenotetext{c}{SIMBAD: Blue straggler; \cite{2011AA...534A.125U}: $\gamma$ Dor; \cite{2013MNRAS.430.3472B}: suspected $\gamma$ Dor; \cite{2020PhDT........26C}: $\gamma$ Dor; \cite{2014AJ....148...38M}: Single-line spectroscopic binary, 762 d orbital period.}
\tablenotetext{d}{\cite{2020PhDT........26C}: $\delta$ Sct star.}
\tablenotetext{e}{SIMBAD: Eclipsing binary; \cite{2010AJ....140.1268T}: Eclipsing binary, 0.3487 d orbital period; \cite{2012ApJ...745...57G}: x-ray active binary, possible RS CVn variable; \cite{2020PhDT........26C}: Possible high-amplitude $\delta$ Sct star.}
\end{deluxetable*}
 
\section{Spectroscopy} \label{sec:spectroscopy}

We did not find stellar parameters derived from spectroscopy for these five stars in the literature, for example, from the LAMOST ROTFIT pipeline \citep{2022arXiv220504757F}. We have taken new low-resolution spectra (R $\sim$ 2000) for three of the stars, KIC\,5024468, KIC\,5113357, and KIC\,5112843, using the ALFOSC spectrograph mounted on the 2.56-meter Nordic Optical Telescope (NOT) at the Roque de los Muchachos Observatory on La Palma.

\begin{deluxetable*}{lccc}
\tablenum{3}
\tablecaption{Summary of spectroscopic results.\label{tab:spectroscopy}}
\tablewidth{0pt}
\tablehead{
\colhead{KIC} &  \colhead{$T_{\rm eff}$ (K)} & \colhead{\loggcms} & \colhead{[M/H]}
}
\startdata
5024468	& 7770 $\pm$ 90 & 4.381 $\pm$ 0.081 & -0.196 $\pm$ 0.306 \\
5113357	&   7270 $\pm$ 90 & 3.709 $\pm$ 0.039 & -0.529 $\pm$ 0.324\\
5112843	&   5600 $\pm$ 150	& 4.37 $\pm$ 0.05 & -1.424 $\pm$ 0.174\\
\enddata
\end{deluxetable*}


The observed spectra were reduced using standard longslit and \'{e}chelle data processing techniques and IRAF packages.
All spectra were modeled with interpolated local thermal equilibrium (LTE) synthetic spectra drawn from the BOSZ \citep{bohlin17} spectral library to determine the fundamental atmospheric parameters. The BOSZ library was calculated for scaled solar metallicity with carbon and $\alpha$-element enhancement; therefore, individual abundance patterns cannot be investigated with our method. Table \ref{tab:spectroscopy} summarizes the results for $T_{\rm eff}$, log g, and [M/H] from the spectral analysis.

Our fitting procedure ({\sc XTgrid}; \citealt{nemeth12}) is based on a steepest-gradient chi-square minimizing method, which was developed to model hot stars. 
To improve its performance for cool stars, we added a grid-search preconditioning to the procedure. 
We step through a set of models to search for the best starting model for the steepest-descent part. 
Next, the descent part takes over in driving the fit and converges on the best solution. 
Once a convergence is achieved, the procedure explores the parameter errors by stepping through a set of points
around the best solution. 
If a better solution is found during error calculations, then the procedure returns to the descent part, hence pushing the solution towards the global minimum. 
{\sc XTgrid} fits the radial velocity and projected rotation velocity of each spectra along with the stellar surface parameters, such as the effective
temperature ($T_{\rm eff}$), surface gravity (log g), and [M/H].

In addition, the procedure acquires photometric data from the VizieR Photometry Viewer\footnote{\url{http://vizier.u-strasbg.fr/vizier/sed/}}, distance data from the Gaia EDR3 database, and extinction values from the NED online services. 
The spectroscopic surface parameters combined
with these measurements allow us to reduce systematics and derive absolute stellar parameters, such as mass, radius, and luminosity. 
An anti-correlation is observed between $T_{\rm eff}$ and [Fe/H]. 
Fortunately, the spectral energy distribution (SED) helps in resolving this bias by restricting the $T_{\rm eff}$. 
Another bias is observed in surface gravity, in particular at low temperature, where the spectrum
is insensitive to the surface gravity. Therefore, the derived log g for KIC 5112843, having the lowest $T_{\rm eff}$, is particularly uncertain.



\section{Steps Toward Asteroseismology for NGC 6819 Blue Stragglers}\label{sec:asteroseismology}

Although we have found a rich spectrum of modes in the {\it Kepler} data for NGC 6819 blue stragglers, and these stars have the added constraints of common age and possibly initial element abundances, asteroseismology for blue stragglers in general, and for these stars, in particular, is difficult in practice.  The pulsation modes in these blue stragglers are not obviously amenable to mode identification.  In addition, modeling of binary interactions and mergers with unknown history requires exploration of many additional parameters.  We summarize some efforts from the literature toward asteroseismology of blue stragglers and take a few first steps toward deriving properties of the NGC 6819 blue straggler stars using asteroseismic techniques.

\subsection{Blue straggler modeling and asteroseismology}

Photometrically variable cluster and field blue stragglers were known before the contributions of space missions such as {\it Kepler}.  Blue straggler asteroseismology attempts have focused on high-amplitude $\delta$ Sct (HADS) or SX Phe stars, for which the radial fundamental and/or first-overtone modes are expected. Mode identification can sometimes be confirmed from period ratios; the expected first-overtone to fundamental period ratio is $\sim$ 0.77 \citep[see, e.g.,][]{1998ApJ...507..818G}.

\cite{1993ASPC...53...74M} reviews known photometrically variable blue stragglers in stellar systems older than 2-3 Gyr.  Mateo estimates masses for cluster Dwarf Cepheid (a.k.a. HADS) blue stragglers, given as a ratio of the blue-straggler mass to that of the cluster RR Lyr stars having similar color.  Mateo also estimates masses of blue stragglers in eclipsing binary systems based on their light curve, and outlines expectations for blue stragglers forming as a result of stellar mergers.

\cite{1998ApJ...507..818G} calculate evolution and linear nonadiabatic pulsation models to derive theoretical relationships for evolutionary and pulsation masses for SX Phe variable and apply these to estimate masses for four double-mode SX Phe blue stragglers in the globular cluster 47 Tuc.

\cite{2002ApJ...576..963T} evolve grids of single-star normal-helium and high-helium models for blue-straggler SX Phe stars with metallicity representative of globular cluster M55. They find that period-luminosity relations are unaffected by blue-straggler formation if blue stragglers are fully mixed stellar mergers.

\cite{2014ApJ...783...34F} use Hubble Space Telescope images to characterize SX Phe variability in the Galactic globular cluster NGC 6541.  They estimate pulsation masses using linear nonadiabatic models, finding good agreement with predictions of single-star evolution tracks.  \cite{2015ApJ...810...15F} calculate a grid of nonlinear radial pulsation models with metallicities representative of SX Phe variables in Galactic globular clusters and dwarf spheroidal galaxies, and use these to investigate the topology of the SX Phe instability strip.

\cite{2007MNRAS.378.1371B} discuss results of a multi-site campaign to identify $\delta$ Sct pulsations in the open cluster M67, and find two blue stragglers, EW Cnc and EX Cnc, with 46 and 21 frequencies, respectively.  They calculate a grid of pulsation models taking into account rotation, and compare frequency predictions with observations, but conclude that further progress cannot be made without mode identification from spectroscopy or multicolor photometry.

\subsection{Blue straggler asteroseismology using {\it Kepler} or {\it TESS} data}

Asteroseismology has been attempted using {\it Kepler} or {\it TESS} data for only a handful of blue stragglers.  \cite{2021ApJ...923..244H} calculate nonstandard models for the field $\delta$ Sct star KIC 11145123, which is a possible blue straggler.  They explore models with artificially modified envelope composition representing the effects of interaction with another star in forming the blue straggler. They find the best fit to observed frequencies for models with enhanced envelope helium abundances.  \cite{2021ApJ...923..244H} make use of rotational frequency splittings seen in $p$, $g$, and mixed modes for this star to constrain mode identification, and point out that only two other main-sequence stars studied using {\it Kepler} data have been found so far that show such well-resolved frequency splittings.

\cite{2016ApJ...832L..13L} and \cite{2018PhDT........47L} use solar-like oscillations found in {\it Kepler K2} data and asteroseismic scaling relations to determine the mass and radius of S1237, a `yellow' stragger in M67 which presumably evolved from a blue straggler.  The derived mass of S1237 is 2.9 $\pm$ 0.2 M$_{\odot}$, more than twice the mass of the main-sequence turnoff stars.

\cite{2019MNRAS.490.4040A} use {\it TESS} data and single-star evolution models to derive stellar parameters of the low-metallicity high-amplitude $\delta$ Sct star SX Phe that is a possible field blue straggler.  The likely 1st-overtone mode frequency was identified by its characteristic ratio with the fundamental mode.  The {\it TESS} data enable frequencies to be measured to high accuracy. The frequency ratio was fit to the 5th decimal place by a model of 1.05 M$_{\odot}$, initial hydrogen mass fraction $X_{o}$=0.667, $Z$=0.002, and age 2.8 Gyr. This initial hydrogen abundance is lower than normally used for single-star evolution models, indicating that SX Phe could have experienced a prior stellar interaction or merger event.

\subsection{Blue straggler formation and evolution modeling}

Many researchers have explored blue straggler formation scenarios, including mass transfer from a companion or binary merger, without consideration of asteroseismic constraints.

\cite{2015ASSL..413..277S} introduces methods to model blue straggler evolution and discusses success of various approaches in describing observations  \cite{1997ApJ...477..335S} use a smooth particle hydrodynamics code to compare the properties of blue stragglers formed by direct collision with those resulting from binary merger and examine subsequent evolution of the merger products, including mass loss.  They conclude that color distribution is an important constraint for blue-straggler formation scenarios, and that observed color distributions appear to rule out fully mixed models. \cite{2019ApJ...885...45G} use the binary modeling capabilities in the MESA \citep{2013ApJS..208....4P, 2015ApJS..220...15P} stellar evolution code to constrain mass-transfer scenarios for two blue straggler + white dwarf binary systems in NGC 188. \cite{2019ApJ...876L..33P} use MESA and other codes to model stellar mergers and investigate the origin of two populations of blue stragglers in the globular cluster M30.  \cite{2019ApJ...876L..33P} suggest that the redder population is a result of continuous formation of blue stragglers during 10 Gyr via mass transfer and mergers, while a bluer population is the result of stellar collisions during cluster core collapse around 3.2 Gyr ago.  

Y. G\"otberg, in a module for the 2021 MESA summer school available at Zenodo,\footnote{doi: 10.5281/zenodo.523, https://zenodo.org/record/5234616\#.Y63zRsHMId0} outlines methods inspired by \cite{2020ApJ...904L..13R} to use MESA binary capabilities \citep{2015ApJS..220...15P} to model mass transfer, merger, envelope ejection, and subsequent stellar evolution.  This treatment should be applicable for generating blue straggler models and subsequently calculating their pulsations properties to compare with observations. 

\subsection{Outline for evolution modeling of NGC 6819 blue stragglers}

While it is beyond the scope of this paper to calculate asteroseismic models for the four NGC 6819 blue stragglers pulsators discussed here, we attempt to outline to scope of the evolution modeling problem.  One procedure, following G\"otberg (2021), would be to:

1) Use the MESA binary module to model the evolution of detached stars in a binary system; initial masses and separations are free parameters; 2) evolve the binary through the mass transfer phase and up to the common envelope phase, at which point the code will stop; 3) estimate how much mass is ejected during merger (another free parameter); 4) decide the entropy profile of the merged component (there are many reasonable choices); 5) calculate the remaining evolution to 2.4 Gyr, the age of NGC 6819; 6) calculate the pulsation frequencies of the resulting models and compare to observations.  

Steps 1 and 2 are done using the MESA binary module \citep{2015ApJS..220...15P}.  For steps 3-5 above, given a number of assumptions, MESA can build the merged relaxed model for further evolution.  Discussion and example MESA inlists are available at either the G\"otberg Zenodo link or at the Zenodo link in the Appendix of \cite{2020ApJ...904L..13R}.  The evolution of the merger product can be continued with MESA (step 5). The model frequencies can be calculated (step 6) with the GYRE in MESA capabilities, or using the GYRE stellar oscillation code \citep{2013MNRAS.435.3406T} separately.

There are of course many choices and settings in MESA to consider, for example initial metallicity and helium abundance, initial stellar rotation rates, convection and convective overshooting treatment, opacities and abundance mixture, etc.

Besides a binary merger, there are other possible formation scenarios for the NGC 6819 blue stragglers.  For example, these stars could have accreted mass from an undetected binary companion, or from a close interaction or collision with a passing star.  There is evidence that at least one of the stars discussed here, KIC 5113357, may show a binary orbital period in its spectrum.  There is also the remote possibility that these blue stragglers are interlopers to the cluster that happen to have the nearly the same kinematic properties as the other cluster members.  In the latter case, it might be appropriate to model the stars using single-star evolution, ignoring the cluster age constraint.  Perhaps such models would help to rule out this scenario. 

\subsection{Applying asteroseismic scaling relations to NGC 6819 {$\delta$} Scuti blue stragglers}

We apply asteroseismic `scaling relations' developed for $\delta$ Sct stars that point to a mean density and effective temperature for KIC 5024468 and KIC 5113357, which can then be used to estimate other stellar properties.  For the two $\gamma$ Dor candidates, KIC 5024084 and KIC 5024455, we search for period spacing sequences, which may hold information on near-core rotation rates.

\subsubsection{Frequency separation--mean-density relation}

Unlike the case for solar-like oscillators, $\delta$ Sct pulsations have low radial order and are not in the asymptotic regime, and therefore do not show uniform frequency separations between modes of the same angular degree $\ell$ and consecutive radial order $n$ as do solar-like oscillators.  Nevertheless, regularities in frequency separations have been found in $\delta$ Sct stars \citep[see, e.g.,][]{2014A&A...563A...7S, 2016ApJ...822..100P, 2016ApJS..224...41P, 2020Natur.581..147B} that could be associated with a large separation.  \cite{2014A&A...563A...7S} developed a frequency separation ($\Delta\nu$)–mean-density relation for $\delta$ Sct stars based on stellar modeling.  \cite{2015ApJ...811L..29G} confirmed this relation observationally using $\delta$ Sct stars in eclipsing binaries:

\begin{equation}\label{eq:Eq1}
\frac{\bar\rho}{\bar\rho_{\odot}} = (1.55^{+1.07}_{-0.68})(\frac{\Delta \nu}{\Delta \nu_{\odot}})^{2.035 \pm 0.095}
\end{equation}

\noindent The relation is normalized to solar values, with $\Delta\nu_{\odot}$ = 134.8 $\mu$Hz (Kjeldsen 2008).   

For the two $\delta$ Sct stars with many significant frequencies, KIC 5024468 and KIC 5113357, we calculated the frequency difference between each mode and all of the other modes and created a histogram of frequency spacings.  This method was used to find characteristic spacings for $\delta$ Sct stars by, e.g., \cite{2008CoAst.157...56B}, \cite{2009MNRAS.396..291B}, and \cite{2011A&A...533A.133Z}.  We excluded from this procedure frequencies less than 3 c/d that are not likely to be $\delta$ Sct modes, and also experimented with thresholds for exclusions of modes below a S/N threshold, which may be higher-degree modes or less-reliable detections.

\begin{figure*}
\gridline{\fig{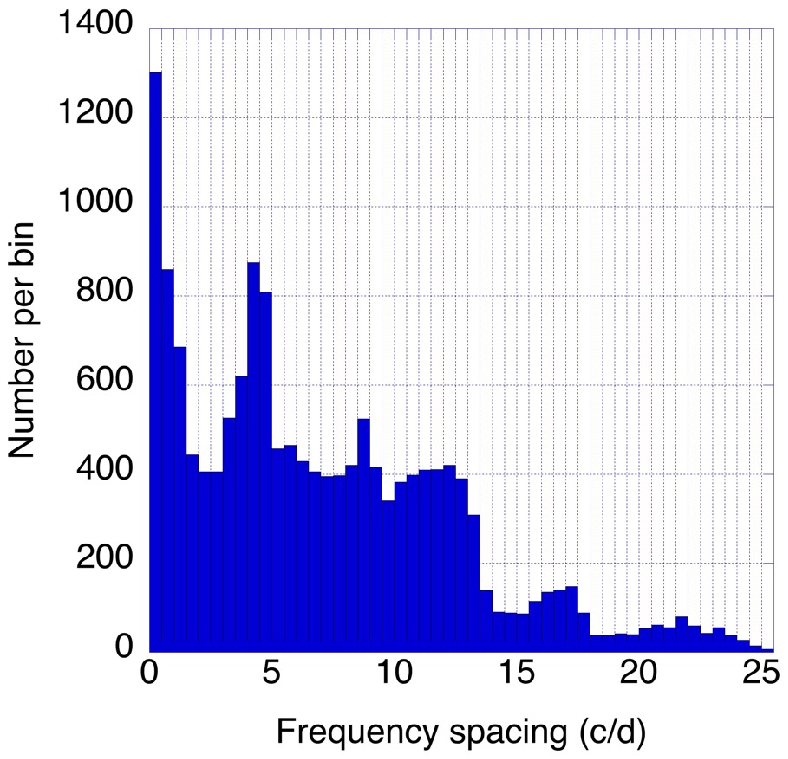}{0.4\textwidth}{(a)}
          \fig{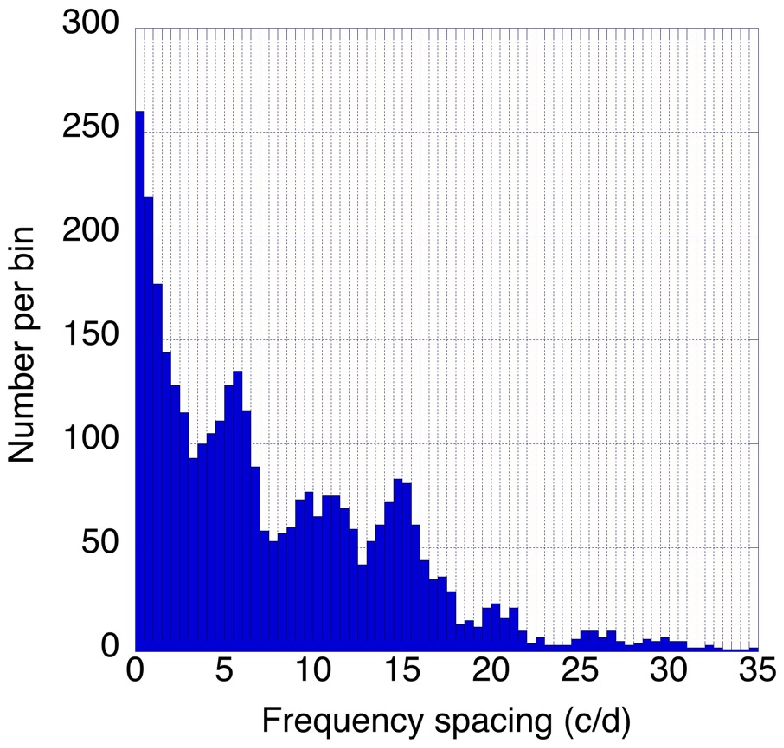}{0.4\textwidth}{(b)}
          }
\caption{Distribution of frequency spacings for $\delta$ Sct stars KIC 5024468 (a) and KIC 5113357 (b).}
\label{fig:freqspachistograms}
\end{figure*}

Figure \ref{fig:freqspachistograms} shows the results for these two stars, excluding modes with S/N $<$ 10. Ignoring the peak near 0 c/d, which is a consequence of modes that are closely spaced that may occur for several reasons (e.g., modes of different angular degree with coincidentally the nearly the same frequency, or rotational splitting), we find peaks at 4.0-4.5 c/d for KIC 5024468, and 5.5-6 c/d for KIC 5113357.

\begin{figure*}
\gridline{\fig{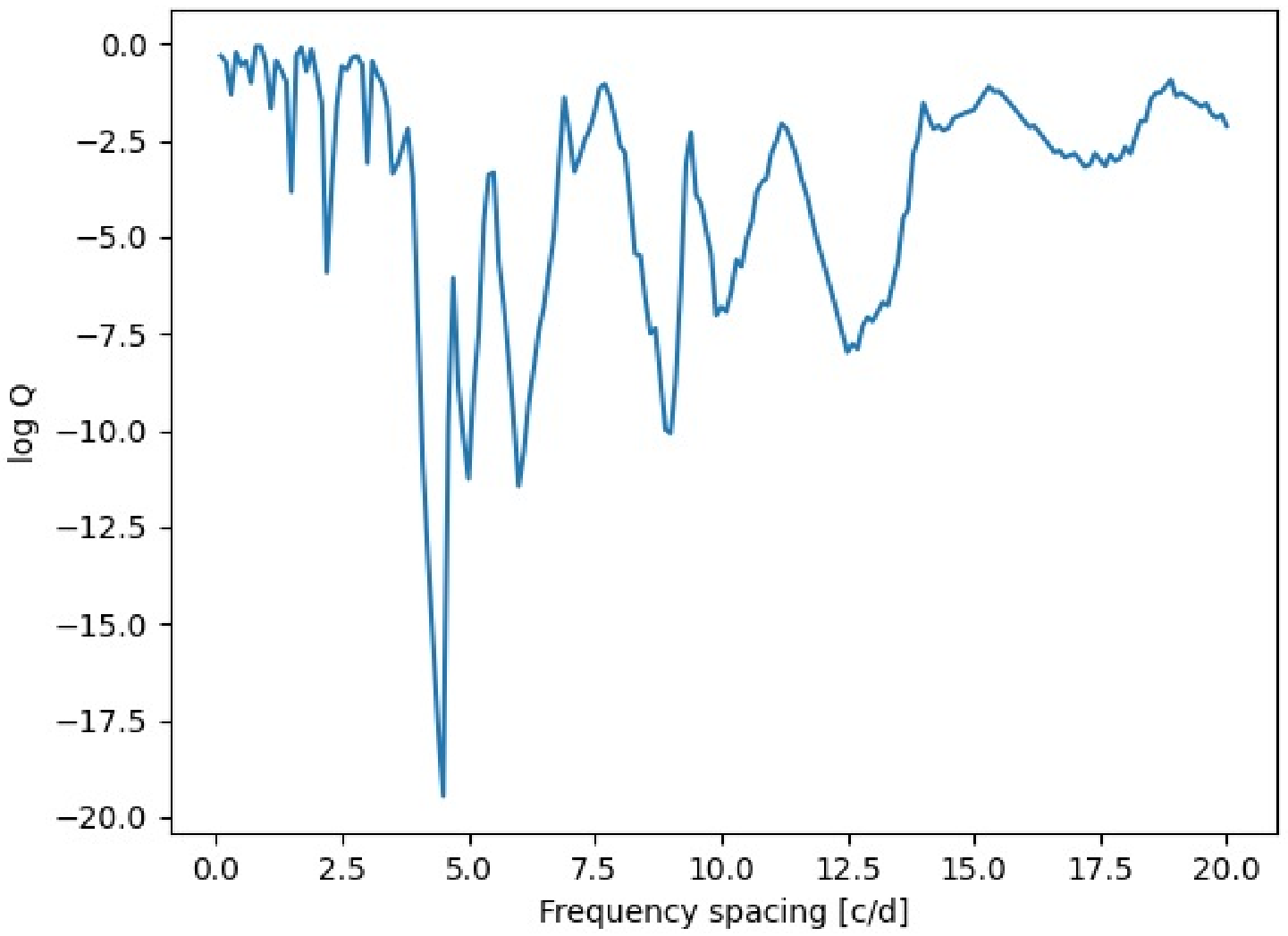}{0.4\textwidth}{(a)}
          \fig{KStestKIC5113357.eps}{0.4\textwidth}{(b)}
          }
\caption{Kolmogorov-Smirnov test results for frequency separations of $\delta$ Sct stars KIC 5024468 (a) and KIC 5113357 (b).}
\label{fig:KStest}
\end{figure*}

To estimate a common frequency spacing among modes, we also applied the Kolmogorov–Smirnov (K-S) test as discussed by \cite{1988IAUS..123..329K} to test the significance of a perceived uniform spacing.  We plotted the results in Figure \ref{fig:KStest}, again excluding frequencies with S/N $<$ 10.  Non-random periods spacings will appear as minima in Q, where Q is the probability that spacings are randomly distributed. The ‘confidence level’ that a given spacing is significant is (1-Q) $\times$ 100\%.  There are minima in these two plots coinciding well with the histogram results, giving 4.4 c/d for KIC 5024468 and 5.5 c/d for KIC 5113357.  Note that for KIC 5113357 the deepest minima occur at around 12 c/d, or around twice the value of the selected frequency spacing value, but physically we expect that the smallest frequency spacing is the actual one, and that multiples of this spacing should also appear in the analysis. Likewise, for KIC 5024468, while the deepest minimum in the K-S test is at 4.4 c/d, there is a shallower but significant minimum around half this value at 2.3 c/d, and so this value is more likely to be the actual frequency spacing corresponding to the mean density. We also do not expect the spacing to be exact for these modes in the non-asymptotic regime, so there should be a spread in frequency separations, which will make the dips shallower.  

Using these frequency spacings in the asteroseismic scaling relation (Equation \ref{eq:Eq1}) results in a mean density of 0.0571 $\bar\rho_{\odot}$ for KIC 5024468 and 0.3367 $\bar\rho_{\odot}$ for KIC 5113357.  Taking into account the uncertainties in the asteroseismic scaling relation, the mean densities of the two stars are 0.0.037-0.083 $\bar\rho_{\odot}$, and 0.203-0.530 $\bar\rho_{\odot}$, respectively.

\vspace{5pt}

\subsubsection{$\bar{T}_{\rm eff}-\nu_{max}$ scaling relation}

\cite{2018A&A...614A..46B} derive a mean $T_{\rm eff}$--$\nu_{max}$ scaling relation for $\delta$ Sct stars based on data from over 1000 stars observed during the {\it CoRoT} \citep{2006ESASP.624E..34B} and {\it Kepler} missions:

\begin{equation}\label{eq:Eq2}
\bar{T}_{\rm eff} (K) = (2.94 \pm 0.24) \nu_{max}(\mu Hz) + (6980 \pm 50)
\end{equation}

The frequencies with maximum amplitude for KIC 5024468 and KIC 5113357 are 147.05 $\mu$Hz and 198.93 $\mu$Hz, respectively, giving $T_{\rm eff}$ values, according to this scaling relation, of 7412 $\pm$ 85 K and 7565 $\pm$ 98 K. For KIC 5024468, the scaling-relation $T_{\rm eff}$ is between the higher $T_{\rm eff}$ obtained from spectroscopy in Table \ref{tab:spectroscopy} and the lower $T_{\rm eff}$ of the {\it TESS} Input Catalog in Table \ref{tab:properties}.  For KIC 5113357, the $T_{\rm eff}$ from the scaling relation is slightly hotter than, but not inconsistent with, the $T_{\rm eff}$ values in Table \ref{tab:properties} and Table \ref{tab:spectroscopy}.

The $T_{\rm eff}$ and mean density derived from these asteroseismic scaling relations can be used in principle to calculate the stellar mass, radius, and luminosity of these two stars.  These estimates can be approached in two ways:

1) First, adopting a log g value, the mass and radius are constrained by the mean density.  Then, using $T_{\rm eff}$, the luminosity can be obtained.  This method is very sensitive to the value of surface gravity, g, because mass is proportional to g$^3$.  

2) Alternatively, assuming the TIC luminosity value (Table \ref{tab:properties}) is reasonably accurate, one could use $T_{\rm eff}$ to derive the radius, and use radius and mean density to derive stellar mass.  This second method does not make use of the surface gravity value.  We can compare the TIC luminosity values with those found in the Gaia DR2 archive \citep{2016A&A...595A...1G, 2018A&A...616A...1G}:  The Gaia DR2 luminosities for KIC 5024468 and KIC 5113357 are 32.3 L$_{\odot}$ and 7.2 L$_{\odot}$, respectively, very close to the TIC values of 34.6 $\pm$ 4.0 L$_{\odot}$ and 8.5 L$_{\odot}$.

The uncertainties in derived mean density and measured log g are large enough that the parameters of the two stars are not well constrained.  Nevertheless, we can adopt the mean density and $T_{\rm eff}$ from the asteroseismic scaling relations and apply both methods, adjusting log g to attain consistency with the TIC luminosity value.  The results of this exercise are summarized in Table \ref{tab:scalingrelations}.

\begin{deluxetable*}{lcc}
\tablenum{4}
\tablecaption{Parameters inferred for NGC 6819 blue straggler $\delta$ Sct stars from $\Delta\nu-\bar\rho$  and $T_{\rm eff}$--$\nu_{max}$ scaling relations. \label{tab:scalingrelations}}
\tablewidth{0pt}
\tablehead{
\colhead{KIC}  & {5024468} & \colhead{5113357} \\
}
\startdata
$\Delta\nu$ (c/d)   &  2.3  &  5.5\\
$\nu_{max}$ ($\mu$Hz)  &   147.05  &  198.93\\
\hline
Mean Density ($\bar\rho_{\odot}$)  &    0.0571  &  0.3367\\
$T_{\rm eff}$ (K) &    7412  &  7565 \\
Adjusted log g  &  3.72 &  4.19 \\
Mass  (M$_{\odot}$) &  2.15  &  1.59\\
Radius  (R$_{\odot}$) & 3.35  &  1.68\\
Resulting Luminosity (L$_{\odot}$)   &  30.4  &  8.28 \\
\enddata
\end{deluxetable*}

For KIC 5024468, the derived luminosity can be made consistent with the TIC luminosity value if a log g= 3.72 is adopted.  This value is slightly higher than that of the TIC, but lower than the value obtained from the analysis of the low-resolution spectroscopy of Table \ref{tab:spectroscopy}. The resulting stellar mass is 2.15 M$_{\odot}$, reasonable for a blue straggler $\delta$ Sct star cluster member.  If we had instead adopted a mean frequency separation $\Delta\nu$ = 4.4 c/d, a high log g value of 4.3, consistent with the value in Table \ref{tab:spectroscopy}, would have been required to attain consistency with the TIC luminosity.   The inferred stellar mass, however, would have been 8.4 M$_{\odot}$, much higher than the estimate in the TIC, and unreasonably high for a $\delta$ Sct star and probably also for a blue straggler cluster member resulting from stellar interactions or a merger.


For KIC 5113357, the derived luminosity can made approximately consistent with the TIC luminosity value if a log g of 4.19 is adopted, which is close to that of the TIC value, but higher than the value from spectroscopy in Table \ref{tab:spectroscopy}.  If the Table \ref{tab:spectroscopy} log g value, 3.7, is adopted instead, the derived stellar mass is unreasonably low, 0.054 M$_{\odot}$.

\cite{2009MNRAS.396..291B} use theoretical models to correlate mean separations of $\delta$ Sct radial modes with log g values.  According to their Figure 8, models with a mean separation of 2.3 c/d have log g = 3.7, and models with a mean separation of 5.5 c/d have log g = 4.2, confirming our log g inferences in Table \ref{tab:scalingrelations}, columns 2 and 3.  However, there is a caveat that the \cite{2009MNRAS.396..291B} results are based on single-star evolution models.  

\subsection{Period spacings for $\gamma$ Doradus stars}

As discussed by, e.g., \cite{2015ApJS..218...27V, 2015A&A...574A..17V, 2016A&A...593A.120V, 2018A&A...618A..24V, 2018MNRAS.474.2774S, 2019MNRAS.482.1757L, 2019MNRAS.487..782L, 2020MNRAS.491.3586L}, gravity-mode and global Rossby-mode period-spacing patterns in $\gamma$ Dor variables can be used to probe internal rotation and even to detect differential rotation. \cite{2020MNRAS.491.3586L} found clear period spacing sequences in 611 out of 2085 $\gamma$ Dor variables observed for four years by {\it Kepler}.  The slope of Rossby-mode period spacing vs.\,period is positive as these modes propagate retrograde to the rotation, while the slope of of prograde and zonal gravity modes is negative. 

\begin{figure*}
\gridline{\fig{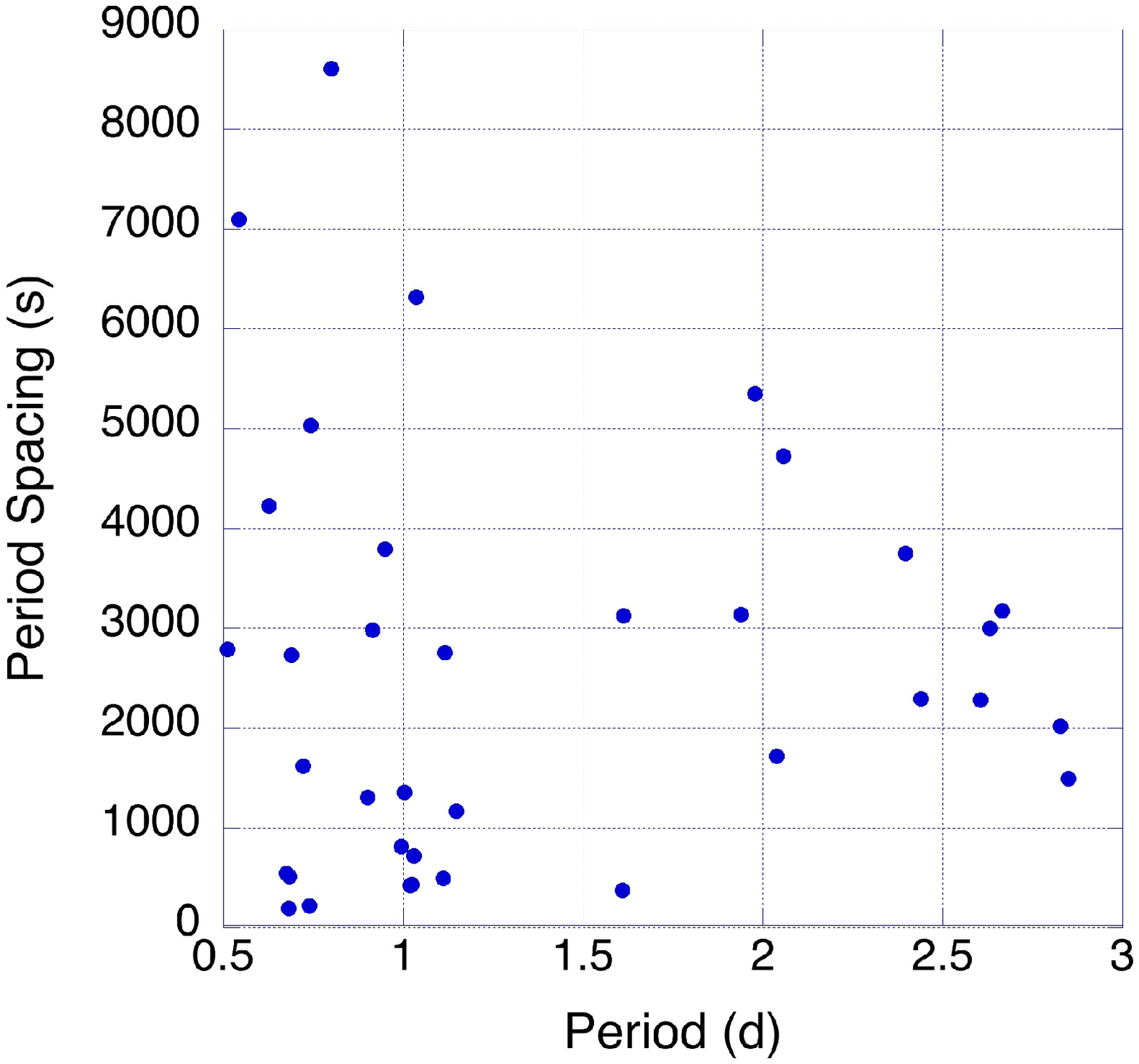}{0.4\textwidth}{(a)}
          \fig{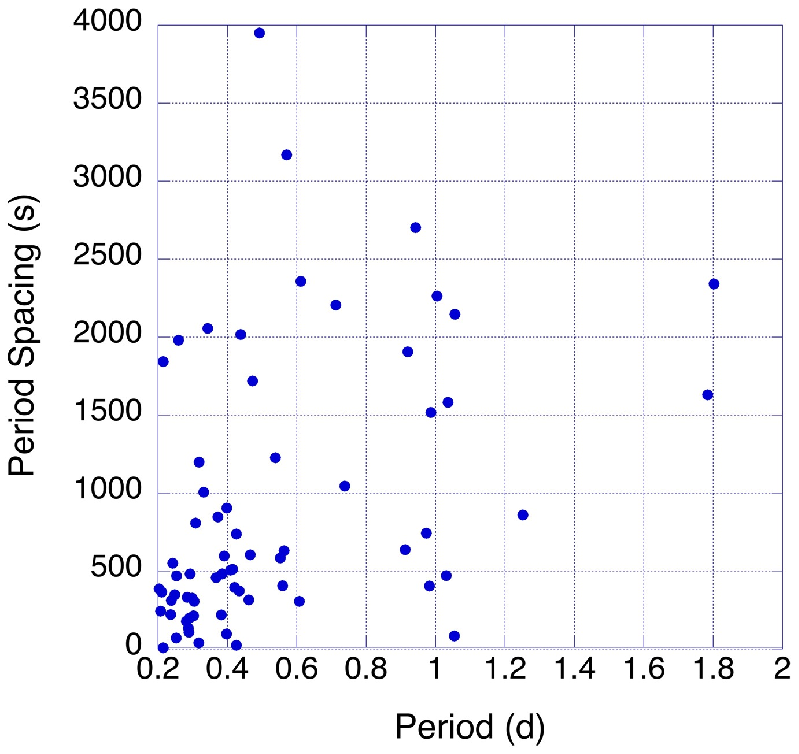}{0.4\textwidth}{(b)}
          }
\caption{Spacing between consecutive periods vs. period for $\gamma$ Dor candidates KIC 5024084 (a) and KIC 5024455 (b).}
\label{fig:periodspacing}
\end{figure*}

We attempted to identify period spacings for the blue straggler $\gamma$ Dor candidates KIC 5024084 and KIC 5024455.  Figure \ref{fig:periodspacing} shows period spacing vs.\,period for consecutive modes for these two stars. The two plots appear by eye as scatter plots, and clear sequences with positive or negative slope are not evident.  Perhaps this result is not unexpected, because \cite{2020MNRAS.491.3586L} were only able to find clear sequences in 30\% of their original sample.  Also, \cite{2020MNRAS.491.3586L} retained frequencies with S/N $>$ 3 in their light-curve analysis, whereas we have retained frequencies with S/N $>$ 25 for KIC 5024084 and S/N $>$ 8.4 for KIC 5024455. In addition, sequences can deviate from a linear fit because of differential rotation or mode trapping, making them more difficult to identify.

\begin{figure*}
\centering
\gridline{\fig{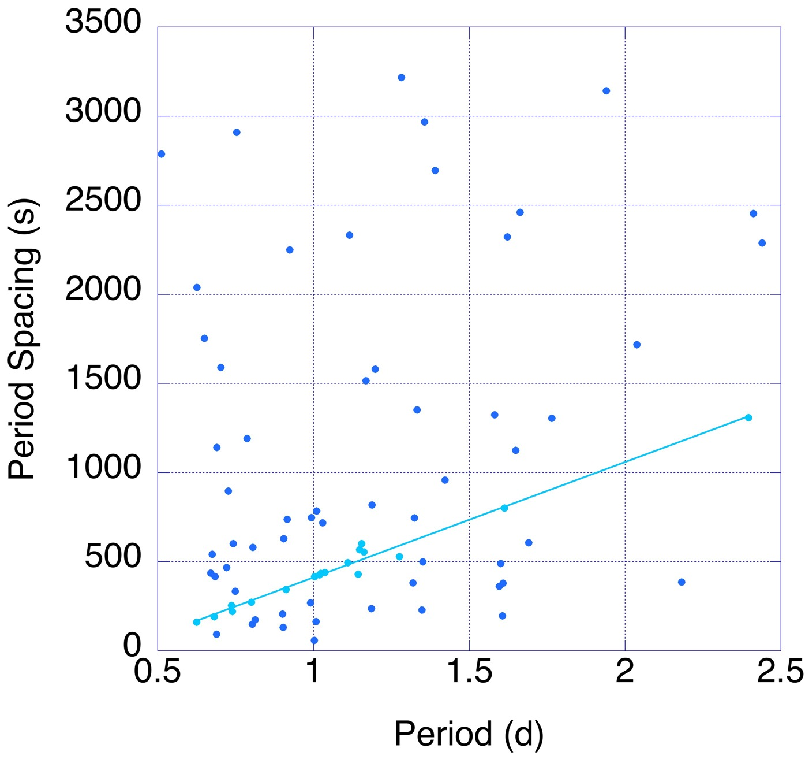}{0.4\textwidth}{(a)}
          \fig{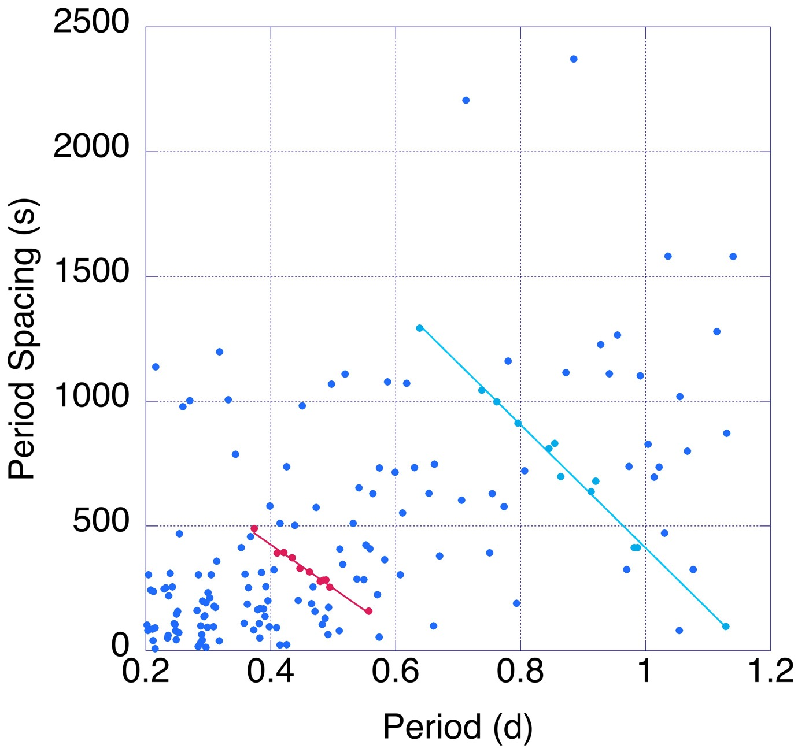}{0.4\textwidth}{(b)}
          }\caption{Spacing between consecutive periods vs.\,period for KIC 5024084 (a) and KIC 5024455 (b) resulting from further pre-whitening of the light curve to identify additional low frequencies. The points in light blue or red represent hypothetical spacing sequences.  The lines show linear regression fits to these sequences.}
\label{fig:newperiodspacing}
\end{figure*}

We ended pre-whitening for low-frequency peaks at S/N $\sim$ 25 for KIC 5024084 and S/N = 8.4 for KIC 5024455 because it became unreliable to pre-whiten additional peaks in the low-frequency region where the noise level increases.  Nevertheless, we attempted further pre-whitening for these two stars to identify additional low-frequency modes. This process increased the number of low-frequency modes from 49 to about 130 for KIC 50244084, and from 84 to 210 for KIC 5024455. Figure \ref{fig:newperiodspacing} shows the resulting period spacings vs.\,period for these two stars, zooming in on the shorter-period region.  No clear sequences are discernible, although there are some sequences of more than 10 points that could represent partial sequences.  Even though this procedure is very subjective, and intended to be illustrative only, we proceeded to identify some hypothetical sequences and consider implications for mode identification and near-core rotation rates.

For KIC 5024084, we chose a hypothetical positive-slope sequence represented by the light-blue points in Figure \ref{fig:newperiodspacing}(a).  This sequence has 18 points; a linear regression fit gives slope = 647.3 s (0.00749 d), $y$-intercept =  -235.34 s, and correlation coefficient R = 0.990.  The mean period of this sequence is 1.115 d, and the average period spacing is 487 s.  Comparing to the values for the 611 stars in Figures 8 and 14 of \cite{2020MNRAS.491.3586L}, these values are similar to those of stars showing $k$ = -2, $m$=-1 Rossby-mode sequences, although the mean spacing is a little low, and the slope a little shallow compared to the stars in the \cite{2020MNRAS.491.3586L} sample.  The implied near-core rotation rate is around 0.5 d$^{-1}$.  Interestingly, this rotation rate coincides with the highest-amplitude frequency suggested as the surface rotation rate by \cite{2013MNRAS.430.3472B}.  \cite{2020MNRAS.491.3586L} were able to identify surface rotation rates for 58 of the 611 stars in their sample, and found that the difference between near-core and surface rotation rates is no larger than 5\%.

For KIC 5024455, we highlighted two hypothetical negative-slope sequences represented by the light-blue and red points in Figure \ref{fig:newperiodspacing} (b).  The first sequence, with 11 points marked in red, has slope =  -1748.4 s (-0.0265 d), $y$-intercept = 1125.2 s, and correlation coefficient R = 0.992.  The mean period of the sequence is 0.459 d, and average period spacing is 323 s.   These values are consistent with an $\ell$=2 $g$-mode sequence according to Figure 8 of \cite{2020MNRAS.491.3586L}.  The second sequence, with 12 points marked in light blue, has slope = -2466.1 s (-0.0285 d), $y$-intercept = 2879 s, and correlation coefficient R = 0.994.  The mean period of the sequence is 0.869 d, and average period spacing is 736 s.  These values are consistent with an $\ell$=1, $m$=1 $g$-mode sequence according to Figures 8 and 14 of \cite{2020MNRAS.491.3586L}, and imply a near-core rotation rate of around 0.6 d$^{-1}$.

\section{Conclusions} \label{sec:conclusions}

Our search for variable stars using pixel data from the {\it Kepler} NGC 6819 superstamp field resulted in identification of five stars that were determined to be cluster members, showing multimode variability in the frequency ranges characteristic of $\gamma$ Dor ($<$ 5 c/d) and/or $\delta$ Sct ($>$5 to $\sim$40 c/d) stars. Two of these stars, KIC 5024468 and KIC 5113357, show a rich spectrum of $\delta$ Scuti pulsation modes, with 236 and 124 significant frequencies identified, respectively, while two stars show mainly low-frequency modes characteristic of either $\gamma$ Dor gravity-mode pulsations or global Rossby modes. Low-frequency variability caused by rotation and star spots cannot be ruled out. The fifth star has an unusual spectrum with several harmonics of two main frequencies. This star shows x-ray activity and may be an RS CVn variable.

We identified frequency separations in the two $\delta$ Sct stars KIC 5024468 and KIC 5113357 that are likely associated with the large frequency spacing $\Delta\nu$ between modes of the same angular degree $\ell$ and consecutive radial order $n$.  Making use of $\delta$ Sct $\Delta\nu-\bar\rho$ and $\bar{T}_{\rm eff}-\nu_{max}$ asteroseismic scaling relations, we were able to estimate mean density and $T_{\rm eff}$. We then used these values in conjunction with log g and luminosity values to estimate stellar masses and radii.

For the two stars showing many low frequencies, KIC 5024084 and KIC 5024455, we could not identify clear sequences of consecutive period spacings as were found by \cite{2020MNRAS.491.3586L} for many $\gamma$ Dor variables.  Nevertheless, we analyzed a hypothetical Rossby-mode sequence for KIC 5024084 and two hypothetical $g$-mode sequences for KIC 5024455, and compared properties of these sequences with those of the Li et al. sample, deriving near-core rotation rates of 0.5 d$^{-1}$ and 0.6 d$^{-1}$, respectively.

We have presented new low-resolution spectroscopic results for only three of the five stars discussed here.  However, some of these spectroscopic results are inconsistent with determinations from asteroseismic inferences for log g  and $T_{\rm eff}$, as well as with the estimates found in the {\it TESS} Input Catalog. Further analysis, as well as additional spectra, particularly high-resolution spectra, would be useful to resolve these discrepancies and to constrain other parameters, for example, stellar rotation rates. Time-series spectra could be examined for radial-velocity variations and used to detect binarity and characterize binary orbits for KIC 5024455, KIC 5113357, and KIC 5112843, possibly providing additional stellar modeling constraints. Time-series spectroscopy and color photometry could also be of use to distinguish frequencies resulting from rotation and star spots from pulsations.

These results from long time-series {\it Kepler} photometry, combined with the common age and element abundances of the cluster members, should provide useful constraints for stellar modeling. These data could be useful for determining the origins and internal structure of the four blue straggler stars.

\begin{acknowledgments}
{\bf Acknowledgments:} We thank the anonymous reviewer for suggestions which greatly improved this paper. We are grateful for data from the NASA {\it Kepler} spacecraft. This research has made use of the SIMBAD database, operated at CDS, Strasbourg, France, and the Mikulski Archive for Space Telescopes (MAST). This work has made use of data from the European Space Agency (ESA) mission {\it Gaia} (\url{https://www.cosmos.esa.int/gaia}), processed by the {\it Gaia} Data Processing and Analysis Consortium (DPAC, \url{https://www.cosmos.esa.int/web/gaia/dpac/consortium}). Funding for the DPAC
has been provided by national institutions, in particular the institutions participating in the {\it Gaia} Multilateral Agreement. J.G. and A.H. acknowledge a Los Alamos National Laboratory Center for Space and Earth Sciences grant CSES XX8P and support from LANL, managed by Triad National Security, LLC for the U.S. DOE’s NNSA, Contract \#89233218CNA000001. J.G. thanks the Society for Astronomical Sciences for the opportunity to present results at their 2022 Symposium, and the American Association of Variable Star Observers for the opportunity to present results at their 110th Annual Meeting. Funding from the National Science Centre in Poland No. UMO-2017/26/E/ST9/00703 and UMO-2017/25/B/ST9/02218 is acknowledged. P.N. acknowledges support from the Grant Agency of the Czech Republic (GA\v{C}R 22-34467S).  The Astronomical Institute in Ond\v{r}ejov is supported by the project RVO:67985815.

\end{acknowledgments}

\pagebreak

\bibliographystyle{aasjournal}



 
\clearpage
\startlongtable
\begin{deluxetable*}{lllllc}
\tablenum{A1}
\tablecaption{Significant frequencies for KIC 5024468.  The harmonics of the 3.05 day eclipsing binary signal of KIC 5024450 are removed from this table.  Real super-Nyquist frequencies are flagged by asterisks.\label{tab:A1}}
\tablewidth{0pt}
\tablehead{
\colhead{Frequency} & \colhead{Frequency} & \colhead{Frequency} & \colhead{Period} & \colhead{Amplitude} & \colhead{S/N } \\
\colhead{number} & \colhead{(c/d)} & \colhead{($\mu$Hz)} & \colhead{(days)} & \colhead{(ppt)} &
}
\startdata
f1 & 12.70528459(26) & 147.051905(3) & 0.0787074064(19) & 10.394(8) & 2779.7 \\ 
f2 & 11.37962770(43) & 131.708654(5) & 0.0878763371(31) & 7.721(8) & 2064.8 \\ 
f3 & 11.9362082(6) & 138.150558(7) & 0.0837786995(42) & 4.761(8) & 1273.3 \\ 
f4 & 10.9057161(13) & 126.223567(15) & 0.091695032(10) & 2.216(8) & 592.7 \\ 
f5 & 11.9883583(39) & 138.754148(45) & 0.083414255(27) & 0.719(8) & 192.4 \\ 
f6 & 10.9727792(43) & 126.99976(5) & 0.091134614(37) & 0.643(8) & 171.8 \\ 
f7 & 12.2835277(43) & 142.17046(5) & 0.081409840(30) & 0.625(8) & 167.1 \\ 
f8 & 13.318839(6) & 154.15324(7) & 0.075081612(35) & 0.494(8) & 132.1 \\ 
f9 & 11.678503(7) & 135.16787(8) & 0.085627403(50) & 0.414(8) & 110.6 \\ 
f10 & 13.315739(8) & 154.11736(9) & 0.075099096(44) & 0.386(8) & 103.2 \\ 
f11 & 12.497896(9) & 144.65158(10) & 0.08001346(6) & 0.331(8) & 88.5 \\ 
f12 & 10.379876(9) & 120.13746(10) & 0.09634025(8) & 0.329(8) & 88.1 \\ 
f13 & 11.020212(10) & 127.54875(11) & 0.09074236(7) & 0.309(8) & 82.7 \\ 
f14 & 24.084919(10) & 278.76064(12) & 0.041519758(17) & 0.278(8) & 74.4 \\ 
f15 & 13.682129(10) & 158.35798(12) & 0.07308803(6) & 0.274(8) & 73.3 \\ 
f16 & 13.703676(11) & 158.60737(13) & 0.07297311(6) & 0.273(8) & 73.1 \\ 
f17 & 13.193131(15) & 152.69828(17) & 0.07579701(8) & 0.261(9) & 69.9 \\ 
f18* & 25.410574(13) & 294.10387(15) & 0.039353695(21) & 0.216(8) & 57.7 \\ 
f19 & 23.976715(16) & 277.50828(18) & 0.041707130(27) & 0.185(8) & 49.5 \\ 
f20 & 13.431526(16) & 155.45748(19) & 0.07445170(9) & 0.178(8) & 47.5 \\ 
f21* & 24.641503(16) & 285.20259(19) & 0.040581938(27) & 0.173(8) & 46.3 \\ 
f22 & 9.804580(16) & 113.47894(19) & 0.10199314(17) & 0.172(8) & 46.0 \\ 
f23 & 8.960029(16) & 103.70405(19) & 0.11160677(21) & 0.171(8) & 45.8 \\ 
f24 & 13.566967(16) & 157.02509(19) & 0.07370843(9) & 0.171(8) & 45.7 \\ 
f25 & 18.128733(16) & 209.82330(19) & 0.055161052(51) & 0.170(8) & 45.6 \\ 
f26 & 21.185620(16) & 245.20394(19) & 0.047201827(37) & 0.168(8) & 45.0 \\ 
f27 & 17.341166(17) & 200.70794(20) & 0.05766624(6) & 0.168(8) & 44.9 \\ 
f28 & 1.325667(17) & 15.34337(20) & 0.754336(9) & 0.165(8) & 44.1 \\ 
f29 & 12.225590(17) & 141.49989(20) & 0.08179563(12) & 0.161(8) & 43.0 \\ 
f30 & 13.187288(18) & 152.63065(21) & 0.07583060(10) & 0.159(8) & 42.7 \\ 
f31 & 18.035824(18) & 208.74797(21) & 0.05544520(6) & 0.158(8) & 42.3 \\ 
f32 & 13.194054(24) & 152.70896(28) & 0.07579171(14) & 0.157(8) & 41.9 \\ 
f33 & 24.219752(18) & 280.32121(21) & 0.041288613(31) & 0.157(8) & 41.9 \\ 
f34 & 9.072612(19) & 105.00709(22) & 0.11022184(23) & 0.152(8) & 40.7 \\ 
f35 & 16.578653(19) & 191.88256(22) & 0.06031853(7) & 0.150(8) & 40.1 \\ 
f36 & 10.463287(19) & 121.10286(22) & 0.09557225(17) & 0.146(8) & 39.1 \\ 
f37 & 13.774664(21) & 159.42899(24) & 0.07259704(10) & 0.139(8) & 37.1 \\ 
f38* & 24.693642(22) & 285.80605(25) & 0.040496252(35) & 0.132(8) & 35.3 \\ 
f39 & 13.057320(22) & 151.12640(26) & 0.07658538(13) & 0.128(8) & 34.2 \\ 
f40 & 13.146638(22) & 152.16017(26) & 0.07606506(13) & 0.126(8) & 33.7 \\ 
f41 & 17.837900(22) & 206.45718(26) & 0.05606040(7) & 0.126(8) & 33.6 \\ 
f42 & 17.727717(22) & 205.18191(26) & 0.05640884(7) & 0.125(8) & 33.4 \\ 
f43 & 10.142700(22) & 117.39237(26) & 0.09859306(22) & 0.125(8) & 33.4 \\ 
f44 & 13.455687(24) & 155.73712(28) & 0.07431802(13) & 0.118(8) & 31.7 \\ 
f45 & 13.614222(25) & 157.57202(29) & 0.07345259(14) & 0.115(8) & 30.7 \\ 
f46 & 12.949320(25) & 149.87640(29) & 0.07722413(15) & 0.115(8) & 30.7 \\ 
f47 & 21.635631(25) & 250.41240(29) & 0.046220052(53) & 0.113(8) & 30.3 \\ 
f48 & 23.962030(25) & 277.33832(29) & 0.041732689(44) & 0.112(8) & 30.0 \\ 
f49 & 13.355554(25) & 154.57818(29) & 0.07487520(14) & 0.112(8) & 29.8 \\ 
f50 & 12.250456(27) & 141.78769(31) & 0.08162960(17) & 0.108(8) & 28.8 \\ 
f51 & 22.759258(27) & 263.41734(31) & 0.043938163(52) & 0.106(8) & 28.4 \\ 
f52 & 23.367969(27) & 270.46261(31) & 0.042793619(49) & 0.106(8) & 28.3 \\ 
f53 & 9.131516(27) & 105.68885(31) & 0.10951083(32) & 0.106(8) & 28.3 \\ 
f54 & 13.049562(28) & 151.03660(32) & 0.07663092(16) & 0.104(8) & 27.7 \\ 
f55 & 6.887328(28) & 79.71445(32) & 0.1451942(6) & 0.101(8) & 27.1 \\ 
f56 & 8.777011(29) & 101.58578(33) & 0.11393399(37) & 0.101(8) & 26.9 \\ 
f57 & 0.310845(29) & 3.59775(34) & 3.21702(30) & 0.100(8) & 26.7 \\ 
f58 & 23.315808(29) & 269.85890(33) & 0.042889355(53) & 0.099(8) & 26.4 \\ 
f59 & 16.823310(29) & 194.71424(33) & 0.05944133(10) & 0.099(8) & 26.4 \\ 
f60 & 17.737015(29) & 205.28953(34) & 0.05637927(9) & 0.098(8) & 26.2 \\ 
f61 & 17.537127(31) & 202.97601(36) & 0.05702188(10) & 0.097(8) & 25.8 \\ 
f62 & 4.816760(30) & 55.74954(35) & 0.2076084(13) & 0.093(8) & 24.9 \\ 
f63 & 21.931019(30) & 253.83124(35) & 0.04559751(6) & 0.092(8) & 24.7 \\ 
f64 & 13.190487(34) & 152.66768(39) & 0.07581221(20) & 0.090(8) & 24.1 \\ 
f65 & 21.288007(31) & 246.38897(36) & 0.04697480(7) & 0.090(8) & 24.1 \\ 
f66 & 13.263203(32) & 153.50930(37) & 0.07539656(19) & 0.090(8) & 24.1 \\ 
f67 & 4.670581(34) & 54.05766(39) & 0.2141060(15) & 0.090(8) & 24.0 \\ 
f68 & 16.449607(32) & 190.38898(37) & 0.06079172(12) & 0.088(8) & 23.7 \\ 
f69 & 13.245851(33) & 153.30847(38) & 0.07549533(19) & 0.088(8) & 23.5 \\ 
f70 & 9.285705(32) & 107.47344(37) & 0.10769240(37) & 0.088(8) & 23.4 \\ 
f71 & 17.888449(33) & 207.04224(38) & 0.05590199(10) & 0.087(8) & 23.2 \\ 
f72 & 11.510335(33) & 133.22147(38) & 0.08687843(24) & 0.086(8) & 23.1 \\ 
f73 & 16.801138(34) & 194.45762(39) & 0.05951978(12) & 0.085(8) & 22.7 \\ 
f74 & 23.924547(34) & 276.90448(39) & 0.04179807(6) & 0.085(8) & 22.7 \\ 
f75 & 13.817980(34) & 159.93033(39) & 0.07236947(17) & 0.085(8) & 22.6 \\ 
f76 & 13.046342(35) & 150.99933(40) & 0.07664983(20) & 0.084(8) & 22.5 \\ 
f77 & 18.031326(35) & 208.69591(40) & 0.05545903(10) & 0.083(8) & 22.2 \\ 
f78 & 4.735994(35) & 54.81475(41) & 0.2111488(16) & 0.082(8) & 21.9 \\ 
f79 & 4.224031(35) & 48.88925(40) & 0.2367406(20) & 0.082(8) & 21.8 \\ 
f80 & 13.164981(35) & 152.37247(41) & 0.07595908(20) & 0.082(8) & 21.8 \\ 
f81 & 22.201270(35) & 256.95915(41) & 0.04504246(7) & 0.080(8) & 21.4 \\ 
f82 & 18.142152(35) & 209.97862(41) & 0.05512025(10) & 0.080(8) & 21.3 \\ 
f83 & 13.628005(36) & 157.73155(42) & 0.07337831(20) & 0.079(8) & 21.0 \\ 
f84 & 23.610974(36) & 273.27517(42) & 0.04235318(7) & 0.078(8) & 20.8 \\ 
f85 & 0.308191(38) & 3.56703(44) & 3.24474(41) & 0.077(8) & 20.5 \\ 
f86 & 12.006543(37) & 138.96462(43) & 0.08328791(25) & 0.076(8) & 20.3 \\ 
f87 & 16.952373(40) & 196.20803(46) & 0.05898878(14) & 0.075(8) & 20.0 \\ 
f88 & 5.434394(38) & 62.89809(44) & 0.1840130(13) & 0.075(8) & 20.0 \\ 
f89 & 13.410414(38) & 155.21313(44) & 0.07456891(21) & 0.075(8) & 20.0 \\ 
f90 & 22.094678(38) & 255.72544(44) & 0.04525976(8) & 0.074(8) & 19.8 \\ 
f91 & 22.554739(39) & 261.05022(45) & 0.04433658(8) & 0.072(8) & 19.3 \\ 
f92 & 22.840653(40) & 264.35941(46) & 0.04378158(8) & 0.072(8) & 19.2 \\ 
f93 & 17.940480(40) & 207.64445(46) & 0.05573986(13) & 0.071(8) & 18.9 \\ 
f94 & 10.714860(41) & 124.01459(47) & 0.09332832(35) & 0.070(8) & 18.7 \\ 
f95 & 23.663162(41) & 273.87920(48) & 0.04225978(7) & 0.069(8) & 18.3 \\ 
f96 & 23.578128(41) & 272.89501(48) & 0.04241218(7) & 0.068(8) & 18.2 \\ 
f97 & 1.030520(42) & 11.92732(49) & 0.970384(39) & 0.067(8) & 17.9 \\ 
f98 & 10.095817(42) & 116.84974(49) & 0.09905091(42) & 0.066(8) & 17.7 \\ 
f99 & 11.214226(43) & 129.79429(50) & 0.08917244(35) & 0.066(8) & 17.5 \\ 
f100 & 17.358701(43) & 200.9109(5) & 0.05760799(14) & 0.065(8) & 17.5 \\ 
f101 & 4.673073(43) & 54.0865(5) & 0.2139920(21) & 0.066(8) & 17.5 \\ 
f102 & 12.581161(43) & 145.6153(5) & 0.07948393(28) & 0.065(8) & 17.3 \\ 
f103 & 13.438889(43) & 155.5427(5) & 0.07441092(24) & 0.064(8) & 17.2 \\ 
f104 & 17.530983(43) & 202.9049(5) & 0.05704187(15) & 0.064(8) & 17.2 \\ 
f105 & 17.575816(43) & 203.4238(5) & 0.05689635(14) & 0.064(8) & 17.0 \\ 
f106 & 11.763100(43) & 136.1470(5) & 0.08501162(32) & 0.063(8) & 16.8 \\ 
f107 & 11.245694(43) & 130.1585(5) & 0.08892293(36) & 0.063(8) & 16.8 \\ 
f108 & 0.556606(43) & 6.4422(5) & 1.79658(15) & 0.062(8) & 16.5 \\ 
f109 & 17.981654(43) & 208.1210(5) & 0.05561223(14) & 0.062(8) & 16.5 \\ 
f110 & 0.452761(43) & 5.2403(5) & 2.20866(22) & 0.061(8) & 16.4 \\ 
f111 & 11.960092(43) & 138.4270(5) & 0.08361136(32) & 0.061(8) & 16.4 \\ 
f112 & 13.701709(52) & 158.5846(6) & 0.07298358(25) & 0.060(8) & 16.2 \\ 
f113 & 13.134674(52) & 152.0217(6) & 0.07613434(28) & 0.060(8) & 16.1 \\ 
f114 & 0.769106(52) & 8.9017(6) & 1.30021(8) & 0.059(8) & 15.8 \\ 
f115 & 17.716587(52) & 205.0531(6) & 0.05644428(16) & 0.058(8) & 15.5 \\ 
f116 & 13.000996(52) & 150.4745(6) & 0.07691718(29) & 0.058(8) & 15.5 \\ 
f117 & 9.256438(52) & 107.1347(6) & 0.1080329(6) & 0.058(8) & 15.5 \\ 
f118 & 22.400038(52) & 259.2597(6) & 0.04464277(10) & 0.057(8) & 15.3 \\ 
f119 & 13.281330(52) & 153.7191(6) & 0.07529369(28) & 0.057(8) & 15.3 \\ 
f120 & 6.811093(52) & 78.8321(6) & 0.1468193(10) & 0.057(8) & 15.3 \\ 
f121 & 12.589179(52) & 145.7081(6) & 0.07943331(31) & 0.057(8) & 15.1 \\ 
f122 & 17.534845(52) & 202.9496(6) & 0.05702930(17) & 0.056(8) & 15.0 \\ 
f123 & 12.317374(52) & 142.5622(6) & 0.08118611(34) & 0.056(8) & 15.0 \\ 
f124 & 18.272433(52) & 211.4865(6) & 0.05472724(15) & 0.055(8) & 14.8 \\ 
f125 & 20.563822(52) & 238.0072(6) & 0.04862908(12) & 0.055(8) & 14.8 \\ 
f126 & 10.237432(52) & 118.4888(6) & 0.09768076(49) & 0.055(8) & 14.8 \\ 
f127 & 18.024802(52) & 208.6204(6) & 0.05547910(16) & 0.055(8) & 14.7 \\ 
f128* & 24.988798(52) & 289.2222(6) & 0.04001793(8) & 0.054(8) & 14.4 \\ 
f129 & 12.257231(52) & 141.8661(6) & 0.08158450(36) & 0.053(8) & 14.1 \\ 
f130 & 0.582603(52) & 6.7431(6) & 1.71642(16) & 0.052(8) & 14.0 \\ 
f131 & 16.67731(6) & 193.0245(7) & 0.05996168(21) & 0.051(8) & 13.8 \\ 
f132 & 21.824130(52) & 252.5941(6) & 0.04582084(12) & 0.051(8) & 13.6 \\ 
f133 & 13.25891(6) & 153.4597(7) & 0.07542094(32) & 0.050(8) & 13.5 \\ 
f134 & 23.05811(6) & 266.8763(7) & 0.04336868(10) & 0.050(8) & 13.4 \\ 
f135 & 17.85129(6) & 206.6122(7) & 0.05601833(17) & 0.050(8) & 13.4 \\ 
f136 & 13.15310(6) & 152.2350(7) & 0.07602766(34) & 0.050(8) & 13.3 \\ 
f137 & 13.58931(6) & 157.2837(7) & 0.07358724(31) & 0.049(8) & 13.1 \\ 
f138 & 16.67957(6) & 193.0506(7) & 0.05995358(22) & 0.049(8) & 13.0 \\ 
f139 & 0.90916(6) & 10.5227(7) & 1.09991(7) & 0.049(8) & 13.0 \\ 
f140 & 12.50865(6) & 144.7761(7) & 0.07994462(38) & 0.048(8) & 12.7 \\ 
f141 & 2.39247(6) & 27.6907(7) & 0.417976(10) & 0.047(8) & 12.6 \\ 
f142 & 20.82103(6) & 240.9842(7) & 0.04802835(14) & 0.047(8) & 12.6 \\ 
f143 & 4.48685(6) & 51.9312(7) & 0.2228732(30) & 0.047(8) & 12.6 \\ 
f144 & 21.56761(6) & 249.6252(7) & 0.04636581(13) & 0.046(8) & 12.4 \\ 
f145 & 22.74192(6) & 263.2167(7) & 0.04397165(12) & 0.046(8) & 12.3 \\ 
f146 & 13.89154(6) & 160.7818(7) & 0.07198621(32) & 0.046(8) & 12.3 \\ 
f147 & 22.11836(7) & 255.9996(8) & 0.04521128(14) & 0.046(8) & 12.2 \\ 
f148 & 13.50152(6) & 156.2677(7) & 0.07406568(34) & 0.046(8) & 12.2 \\ 
f149 & 16.95457(7) & 196.2335(8) & 0.05898112(23) & 0.045(8) & 12.0 \\ 
f150 & 0.87959(6) & 10.1805(7) & 1.13688(8) & 0.045(8) & 11.9 \\ 
f151 & 10.82320(6) & 125.2686(7) & 0.09239403(54) & 0.044(8) & 11.8 \\ 
f152 & 13.13892(7) & 152.0709(8) & 0.07610972(38) & 0.044(8) & 11.8 \\ 
f153 & 22.53944(7) & 260.8732(8) & 0.04436666(13) & 0.043(8) & 11.6 \\ 
f154 & 10.68681(7) & 123.6900(8) & 0.0935732(6) & 0.043(8) & 11.6 \\ 
f155 & 17.24009(7) & 199.5381(8) & 0.05800432(22) & 0.043(8) & 11.6 \\ 
f156 & 22.45057(7) & 259.8446(8) & 0.04454229(13) & 0.043(8) & 11.5 \\ 
f157 & 9.95404(7) & 115.2088(8) & 0.1004616(7) & 0.043(8) & 11.5 \\ 
f158 & 13.17899(7) & 152.5347(8) & 0.07587831(38) & 0.043(8) & 11.5 \\ 
f159 & 10.67286(7) & 123.5285(8) & 0.0936956(6) & 0.043(8) & 11.4 \\ 
f160 & 17.87839(7) & 206.9259(8) & 0.05593343(21) & 0.043(8) & 11.4 \\ 
f161 & 24.38378(7) & 282.2197(8) & 0.04101085(12) & 0.042(8) & 11.4 \\ 
f162 & 13.68631(7) & 158.4064(8) & 0.07306571(36) & 0.042(8) & 11.3 \\ 
f163 & 9.31275(7) & 107.7865(8) & 0.1073796(8) & 0.042(8) & 11.3 \\ 
f164 & 22.69433(7) & 262.6659(8) & 0.04406385(13) & 0.042(8) & 11.2 \\ 
f165 & 16.89863(7) & 195.5861(8) & 0.05917636(23) & 0.042(8) & 11.2 \\ 
f166 & 1.79952(7) & 20.8278(8) & 0.555702(21) & 0.042(8) & 11.1 \\ 
f167 & 16.73948(7) & 193.7440(8) & 0.05973899(24) & 0.041(8) & 11.1 \\ 
f168 & 18.00493(7) & 208.3905(8) & 0.05554032(21) & 0.041(8) & 11.0 \\ 
f169* & 25.63993(7) & 296.7585(8) & 0.03900166(10) & 0.040(8) & 10.7 \\ 
f170 & 22.41573(7) & 259.4414(8) & 0.04461152(15) & 0.040(8) & 10.6 \\ 
f171 & 17.39149(7) & 201.2904(8) & 0.05749937(24) & 0.039(8) & 10.5 \\ 
f172 & 9.64243(7) & 111.6022(8) & 0.1037083(8) & 0.039(8) & 10.5 \\ 
f173 & 17.25998(8) & 199.7684(9) & 0.05793745(25) & 0.039(8) & 10.4 \\ 
f174 & 17.78656(8) & 205.8630(9) & 0.05622222(23) & 0.038(8) & 10.3 \\ 
f175 & 17.84611(8) & 206.5522(9) & 0.05603462(23) & 0.038(8) & 10.3 \\ 
f176 & 1.55634(8) & 18.0133(9) & 0.642528(30) & 0.038(8) & 10.2 \\ 
f177* & 24.93303(8) & 288.5768(9) & 0.04010741(12) & 0.038(8) & 10.2 \\ 
f178 & 16.45431(8) & 190.4435(9) & 0.06077432(28) & 0.037(8) & 9.9 \\ 
f179 & 17.79246(8) & 205.9313(9) & 0.05620356(24) & 0.037(8) & 9.8 \\ 
f180 & 13.75939(8) & 159.2522(9) & 0.07267763(42) & 0.037(8) & 9.8 \\ 
f181 & 20.89842(8) & 241.8799(9) & 0.04785049(17) & 0.036(8) & 9.7 \\ 
f182 & 5.68766(8) & 65.8295(9) & 0.1758188(24) & 0.036(8) & 9.6 \\ 
f183 & 17.09837(8) & 197.8979(9) & 0.05848506(27) & 0.036(8) & 9.6 \\ 
f184 & 18.84215(8) & 218.0805(9) & 0.05307247(22) & 0.036(8) & 9.5 \\ 
f185 & 9.94720(8) & 115.1297(9) & 0.1005307(8) & 0.036(8) & 9.5 \\ 
f186 & 13.29390(8) & 153.8646(9) & 0.07522248(45) & 0.035(8) & 9.4 \\ 
f187 & 21.84898(8) & 252.8818(9) & 0.04576871(17) & 0.035(8) & 9.3 \\ 
f188 & 13.75313(9) & 159.1798(10) & 0.07271071(44) & 0.035(8) & 9.3 \\ 
f189 & 22.41927(9) & 259.4823(10) & 0.04460447(17) & 0.034(8) & 9.2 \\ 
f190 & 17.51061(9) & 202.6692(10) & 0.05710821(27) & 0.035(8) & 9.2 \\ 
f191 & 17.71441(9) & 205.0279(10) & 0.05645120(28) & 0.034(8) & 9.2 \\ 
f192 & 4.52164(9) & 52.3338(10) & 0.2211586(41) & 0.034(8) & 9.2 \\ 
f193 & 3.94422(9) & 45.6508(10) & 0.2535347(53) & 0.034(8) & 9.2 \\ 
f194 & 22.28535(9) & 257.9324(10) & 0.04487251(17) & 0.034(8) & 9.0 \\ 
f195 & 23.18924(9) & 268.3940(10) & 0.04312343(16) & 0.033(8) & 8.9 \\ 
f196 & 5.02746(9) & 58.1882(10) & 0.1989076(35) & 0.033(8) & 8.9 \\ 
f197 & 17.40600(9) & 201.4584(10) & 0.05745142(28) & 0.033(8) & 8.9 \\ 
f198 & 8.72618(9) & 100.9975(10) & 0.1145975(12) & 0.033(8) & 8.9 \\ 
f199 & 8.80859(9) & 101.9513(10) & 0.1135254(12) & 0.033(8) & 8.8 \\ 
f200 & 21.32763(9) & 246.8476(10) & 0.04688752(19) & 0.033(8) & 8.7 \\ 
f201 & 17.26209(10) & 199.7928(11) & 0.05793037(31) & 0.032(8) & 8.7 \\ 
f202 & 4.23451(9) & 49.0106(10) & 0.2361543(49) & 0.032(8) & 8.6 \\ 
f203 & 4.37602(9) & 50.6484(10) & 0.2285180(47) & 0.031(8) & 8.4 \\ 
f204 & 22.38891(9) & 259.1310(10) & 0.04466495(19) & 0.031(8) & 8.4 \\ 
f205 & 17.49727(10) & 202.5148(11) & 0.05715174(30) & 0.031(8) & 8.3 \\ 
f206 & 4.90395(10) & 56.7587(11) & 0.2039170(38) & 0.031(8) & 8.3 \\ 
f207 & 17.67968(10) & 204.6260(11) & 0.05656209(31) & 0.031(8) & 8.3 \\ 
f208 & 22.24395(10) & 257.4532(11) & 0.04495604(19) & 0.031(8) & 8.2 \\ 
f209 & 18.28300(10) & 211.6088(11) & 0.05469562(28) & 0.031(8) & 8.2 \\ 
f210 & 18.89005(10) & 218.6349(11) & 0.05293790(27) & 0.030(8) & 8.1 \\ 
f211 & 18.73677(10) & 216.8608(11) & 0.05337099(27) & 0.030(8) & 8.1 \\ 
f212 & 17.13383(10) & 198.3083(11) & 0.05836403(31) & 0.030(8) & 8.1 \\ 
f213 & 22.14983(10) & 256.3638(11) & 0.04514707(20) & 0.030(8) & 8.1 \\ 
f214 & 4.87829(10) & 56.4617(11) & 0.2049898(39) & 0.030(8) & 8.1 \\ 
f215 & 17.68090(10) & 204.6401(11) & 0.05655820(31) & 0.030(8) & 8.1 \\ 
f216 & 17.70270(10) & 204.8924(11) & 0.05648854(30) & 0.030(8) & 8.0 \\ 
f217 & 22.11951(10) & 256.0129(12) & 0.04520894(22) & 0.029(8) & 7.8 \\ 
f218 & 20.12904(10) & 232.9750(11) & 0.04967945(24) & 0.029(8) & 7.7 \\ 
f219 & 17.86915(10) & 206.8189(12) & 0.05596236(31) & 0.029(8) & 7.6 \\ 
f220 & 6.10560(10) & 70.6667(12) & 0.1637839(27) & 0.028(8) & 7.5 \\ 
f221 & 22.84767(10) & 264.4407(12) & 0.04376812(20) & 0.028(8) & 7.5 \\ 
f222* & 24.73206(10) & 286.2508(12) & 0.04043333(16) & 0.028(8) & 7.4 \\ 
f223 & 18.44271(10) & 213.4574(12) & 0.05422193(30) & 0.028(8) & 7.4 \\ 
f224 & 23.90379(10) & 276.6643(12) & 0.04183436(17) & 0.028(8) & 7.4 \\ 
f225 & 18.60231(10) & 215.3046(12) & 0.05375673(30) & 0.028(8) & 7.4 \\ 
f226 & 8.13258(10) & 94.1271(12) & 0.1229621(16) & 0.027(8) & 7.3 \\ 
f227 & 5.03029(11) & 58.2210(13) & 0.1987954(43) & 0.027(8) & 7.2 \\ 
f228 & 19.34762(10) & 223.9308(12) & 0.05168593(29) & 0.027(8) & 7.1 \\ 
f229 & 17.56050(11) & 203.2466(13) & 0.05694597(36) & 0.025(8) & 6.8 \\ 
f230 & 4.74184(11) & 54.8825(13) & 0.2108883(50) & 0.025(8) & 6.8 \\ 
f231 & 22.89404(11) & 264.9774(13) & 0.04367947(22) & 0.025(8) & 6.7 \\ 
f232 & 18.19355(11) & 210.5735(13) & 0.05496453(35) & 0.025(8) & 6.6 \\ 
f233 & 22.91502(11) & 265.2202(13) & 0.04363949(22) & 0.024(8) & 6.5 \\ 
f234 & 22.87365(11) & 264.7414(13) & 0.04371841(22) & 0.024(8) & 6.5 \\ 
f235 & 18.09123(12) & 209.3893(14) & 0.05527538(36) & 0.024(8) & 6.4 \\ 
f236 & 6.95061(13) & 80.4469(15) & 0.1438721(27) & 0.022(8) & 6.0 \\ 
\enddata
\end{deluxetable*}

\startlongtable
\begin{deluxetable*}{lllllc}
\tablenum{A2}
\tablecaption{Significant frequencies for KIC 5024084.\label{tab:A2}}
\tablewidth{0pt}
\tablehead{
\colhead{Frequency} & \colhead{Frequency} & \colhead{Frequency} & \colhead{Period} & \colhead{Amplitude} & \colhead{S/N } \\
\colhead{number} & \colhead{(c/d)} & \colhead{($\mu$Hz)} & \colhead{(days)} & \colhead{(ppt)} &
}
\startdata
f1 & 0.4905974(9) & 5.678211(10) & 2.03833(37) & 10.599(25) & 2119.7 \\
f2 & 0.4858615(10) & 5.623397(12) & 2.05820(43) & 9.025(25) & 1804.9 \\
f3 & 0.3701680(14) & 4.284352(16) & 2.70148(11) & 6.427(26) & 1285.4 \\
f4 & 0.4732882(32) & 5.477873(37) & 2.112877(14) & 2.899(25) & 579.9 \\
f5 & 0.9764773(35) & 11.301821(41) & 1.02409(37) & 2.648(25) & 529.6 \\
f6 & 0.4098561(37) & 4.743705(43) & 2.439881(22) & 2.483(25) & 496.5 \\
f7 & 0.6198850(41) & 7.174595(47) & 1.61320(11) & 2.418(26) & 483.5 \\
f8 & 0.3802255(46) & 4.40076(5) & 2.630019(31) & 2.017(25) & 403.4 \\
f9 & 0.8607462(47) & 9.96234(5) & 1.161782(58) & 1.940(25) & 388.0 \\
f10 & 0.3487965(50) & 4.03700(6) & 2.867001(42) & 1.922(26) & 384.3 \\
f11 & 0.1156809(54) & 1.33890(6) & 8.64447(41) & 1.673(25) & 334.6 \\
f12 & 0.3509096(59) & 4.06145(7) & 2.849737(48) & 1.661(26) & 332.3 \\
f13 & 0.5618414(58) & 6.50279(7) & 1.779862(19) & 1.570(25) & 314.0 \\
f14 & 0.353813(6) & 4.09506(7) & 2.826351(51) & 1.472(27) & 294.3 \\
f15 & 0.505959(7) & 5.85601(8) & 1.976444(26) & 1.415(25) & 283.0 \\
f16 & 0.251242(7) & 2.90790(8) & 3.98021(11) & 1.274(25) & 254.9 \\
f17 & 0.515432(9) & 5.96565(10) & 1.940118(34) & 1.021(25) & 204.3 \\
f18 & 0.375272(9) & 4.34343(11) & 2.66473(6) & 1.013(26) & 202.7 \\
f19 & 0.621557(10) & 7.19395(12) & 1.608862(27) & 0.953(26) & 190.6 \\
f20 & 0.135640(10) & 1.56990(11) & 7.37248(53) & 0.938(25) & 187.6 \\
f21 & 0.417285(10) & 4.82969(12) & 2.396442(58) & 0.920(25) & 184.1 \\
f22 & 0.405446(10) & 4.69266(12) & 2.46642(6) & 0.906(25) & 181.1 \\
f23 & 0.335499(11) & 3.88310(13) & 2.98063(10) & 0.852(25) & 170.3 \\
f24 & 0.384081(11) & 4.44539(13) & 2.60362(8) & 0.833(25) & 166.6 \\
f25 & 0.076008(11) & 0.87973(13) & 13.1565(19) & 0.824(25) & 164.8 \\
f26 & 0.971701(12) & 11.24654(13) & 1.029123(13) & 0.796(25) & 159.3 \\
f27 & 1.110489(12) & 12.85289(13) & 0.90050(9) & 0.792(25) & 158.3 \\
f28 & 0.981199(12) & 11.35647(14) & 1.019161(13) & 0.770(25) & 154.0 \\
f29 & 0.963903(13) & 11.15628(15) & 1.037449(14) & 0.709(25) & 141.7 \\
f30 & 0.870897(13) & 10.07983(15) & 1.148241(17) & 0.685(25) & 136.9 \\
f31 & 0.606286(14) & 7.01720(16) & 1.649385(37) & 0.678(25) & 135.7 \\
f32 & 0.900409(14) & 10.42141(16) & 1.110605(17) & 0.656(25) & 131.2 \\
f33 & 0.757970(14) & 8.77281(17) & 1.319313(26) & 0.636(25) & 127.3 \\
f34 & 0.895801(16) & 10.36806(18) & 1.116319(20) & 0.589(25) & 117.8 \\
f35 & 1.462341(16) & 16.92525(19) & 0.68383(7) & 0.570(25) & 114.1 \\
f36 & 1.346645(16) & 15.58617(19) & 0.74259(9) & 0.568(25) & 113.6 \\
f37 & 1.052435(19) & 12.18096(22) & 0.950178(17) & 0.469(25) & 93.9 \\
f38 & 1.467073(20) & 16.98001(23) & 0.68163(9) & 0.465(25) & 93.0 \\
f39 & 0.996555(23) & 11.53421(27) & 1.003456(23) & 0.391(25) & 78.1 \\
f40 & 1.248672(24) & 14.45222(28) & 0.800851(15) & 0.377(25) & 75.3 \\
f41 & 1.005964(24) & 11.64310(28) & 0.994072(24) & 0.376(25) & 75.1 \\
f42 & 1.351269(26) & 15.63969(30) & 0.740045(14) & 0.351(25) & 70.2 \\
f43 & 1.386334(29) & 16.04553(34) & 0.721328(15) & 0.313(25) & 62.6 \\
f44 & 1.092107(32) & 12.64013(37) & 0.915661(27) & 0.283(25) & 56.5 \\
f45 & 1.449818(37) & 16.78030(43) & 0.689742(17) & 0.248(25) & 49.5 \\
f46 & 1.480702(39) & 17.13776(45) & 0.675355(17) & 0.232(25) & 46.5 \\
f47 & 1.596343(49) & 18.4762(6) & 0.626432(20) & 0.186(25) & 37.1 \\
f48 & 1.837197(55) & 21.2639(6) & 0.544308(16) & 0.165(25) & 32.9 \\
f49 & 1.952979(56) & 22.6039(6) & 0.512038(15) & 0.163(25) & 32.7 \\
f50 & 11.19904(7) & 129.6185(8) & 0.08929(57) & 0.127(25) & 25.4 \\
f51 & 8.20914(25) & 95.0132(29) & 0.12182(37) & 0.036(25) & 7.3 \\
f52 & 10.61011(29) & 122.8023(34) & 0.09425(25) & 0.031(25) & 6.2 \\
f53 & 11.15403(37) & 129.0976(43) & 0.08965(30) & 0.025(25) & 4.9 \\
\enddata
\end{deluxetable*}

\clearpage
\startlongtable
\begin{deluxetable*}{lllllc}
\tablenum{A3}
\tablecaption{Significant frequencies for KIC 5024455.\label{tab:A3}}
\tablewidth{0pt}
\tablehead{
\colhead{Frequency} & \colhead{Frequency} & \colhead{Frequency} & \colhead{Period} & \colhead{Amplitude} & \colhead{S/N } \\
\colhead{number} & \colhead{(c/d)} & \colhead{($\mu$Hz)} & \colhead{(days)} & \colhead{(ppt)} &
}
\startdata
f1 & 1.3330071(32) & 15.428324(37) & 0.7501834(19) & 1.245(9) & 252.8 \\ 
f2 & 2.2806438(33) & 26.396341(38) & 0.4384726(6) & 1.191(9) & 241.8 \\ 
f3 & 0.8632677(35) & 9.991525(41) & 1.1583891(47) & 1.133(9) & 230.0 \\ 
f4 & 0.5546473(43) & 6.41953(5) & 1.802947(15) & 0.882(9) & 179.0 \\ 
f5 & 1.3548306(52) & 15.68091(6) & 0.7380997(27) & 0.800(9) & 162.4 \\ 
f6 & 2.6147033(52) & 30.26277(6) & 0.3824525(8) & 0.712(9) & 144.5 \\ 
f7 & 0.947651(8) & 10.96819(9) & 1.055239(9) & 0.638(9) & 129.5 \\ 
f8 & 2.509016(7) & 29.03954(8) & 0.3985625(12) & 0.585(9) & 118.7 \\
f9 & 1.182879(7) & 13.69073(8) & 0.8453947(51) & 0.541(9) & 109.8 \\ 
f10 & 0.089178(8) & 1.03216(9) & 11.2134(9) & 0.537(9) & 109.1 \\ 
f11 & 2.303269(7) & 26.65821(8) & 0.4341653(14) & 0.537(9) & 109.0 \\ 
f12 & 2.351040(12) & 27.21112(14) & 0.4253435(23) & 0.455(9) & 92.4 \\ 
f13 & 2.165425(9) & 25.06280(10) & 0.4618030(19) & 0.447(9) & 90.8 \\ 
f14 & 1.061296(10) & 12.28353(11) & 0.942243(8) & 0.406(9) & 82.5 \\ 
f15 & 2.560476(10) & 29.63515(11) & 0.3905523(15) & 0.405(9) & 82.2 \\ 
f16 & 0.925857(10) & 10.71594(12) & 1.080079(12) & 0.397(9) & 80.5 \\ 
f17 & 1.810904(10) & 20.95954(11) & 0.5522103(30) & 0.396(9) & 80.4 \\ 
f18 & 1.095502(10) & 12.67943(12) & 0.912822(8) & 0.379(9) & 76.9 \\ 
f19 & 0.546434(11) & 6.32447(13) & 1.830046(37) & 0.378(9) & 76.7 \\ 
f20 & 2.116692(11) & 24.49875(13) & 0.4724351(24) & 0.357(9) & 72.6 \\ 
f21 & 2.376487(11) & 27.50564(13) & 0.4207891(21) & 0.339(9) & 68.7 \\ 
f22 & 1.751414(12) & 20.27100(14) & 0.5709671(38) & 0.333(9) & 67.5 \\ 
f23 & 1.086734(12) & 12.57794(14) & 0.920188(10) & 0.332(9) & 67.4 \\ 
f24 & 2.721999(12) & 31.50462(14) & 0.3673769(16) & 0.324(9) & 65.7 \\ 
f25 & 0.948519(16) & 10.97823(19) & 1.054275(19) & 0.315(9) & 63.9 \\ 
f26 & 0.330086(13) & 3.82045(15) & 3.02950(12) & 0.303(9) & 61.6 \\ 
f27 & 1.645719(14) & 19.04768(16) & 0.6076370(50) & 0.300(9) & 60.9 \\ 
f28 & 0.265259(14) & 3.07013(16) & 3.76989(20) & 0.286(9) & 58.2 \\ 
f29 & 2.597493(14) & 30.06358(16) & 0.3849865(21) & 0.280(9) & 56.9 \\ 
f30 & 1.027200(15) & 11.88890(17) & 0.973519(14) & 0.279(9) & 56.7 \\ 
f31 & 2.148434(15) & 24.86614(17) & 0.4654550(31) & 0.273(9) & 55.4 \\ 
f32 & 0.159539(15) & 1.84652(17) & 6.2680(6) & 0.272(9) & 55.2 \\ 
f33 & 0.970366(16) & 11.23109(18) & 1.030538(16) & 0.265(9) & 53.8 \\ 
f34 & 0.105700(15) & 1.22339(17) & 9.4606(14) & 0.265(9) & 53.7 \\ 
f35 & 0.131383(15) & 1.52064(17) & 7.6113(9) & 0.260(9) & 52.8 \\ 
f36 & 0.560508(16) & 6.48737(18) & 1.784092(50) & 0.258(9) & 52.5 \\ 
f37 & 0.297833(15) & 3.44715(17) & 3.35756(17) & 0.258(9) & 52.3 \\ 
f38 & 0.995676(16) & 11.52403(18) & 1.004342(16) & 0.249(9) & 50.5 \\ 
f39 & 1.403361(16) & 16.24261(18) & 0.712575(8) & 0.248(9) & 50.3 \\ 
f40 & 2.349474(23) & 27.19299(27) & 0.4256271(42) & 0.246(9) & 49.9 \\ 
f41 & 1.013388(17) & 11.72903(20) & 0.986789(16) & 0.243(9) & 49.4 \\ 
f42 & 2.516044(17) & 29.12089(20) & 0.3974491(28) & 0.232(9) & 47.1 \\ 
f43 & 2.410387(19) & 27.89800(22) & 0.4148710(32) & 0.206(9) & 41.8 \\ 
f44 & 3.413619(19) & 39.50948(22) & 0.2929442(16) & 0.205(9) & 41.7 \\ 
f45 & 1.018195(20) & 11.78467(23) & 0.982129(20) & 0.205(9) & 41.7 \\ 
f46 & 0.798995(20) & 9.24763(23) & 1.251572(31) & 0.198(9) & 40.2 \\ 
f47 & 1.566327(20) & 18.12879(23) & 0.638436(8) & 0.192(9) & 39.0 \\ 
f48 & 3.307999(21) & 38.28703(24) & 0.3022974(19) & 0.189(9) & 38.4 \\ 
f49 & 1.788994(21) & 20.70595(24) & 0.558973(7) & 0.188(9) & 38.2 \\ 
f50 & 0.662505(21) & 7.66789(24) & 1.509420(49) & 0.185(9) & 37.5 \\ 
f51 & 0.169836(22) & 1.96570(25) & 5.8879(8) & 0.185(9) & 37.5 \\ 
f52 & 0.965272(23) & 11.17213(27) & 1.035978(24) & 0.181(9) & 36.6 \\ 
f53 & 1.858616(23) & 21.51176(27) & 0.538034(7) & 0.168(9) & 34.1 \\ 
f54 & 0.212950(24) & 2.46470(28) & 4.69593(53) & 0.164(9) & 33.3 \\ 
f55 & 1.774098(24) & 20.53355(28) & 0.563666(8) & 0.162(9) & 33.0 \\ 
f56 & 2.444861(24) & 28.29701(28) & 0.4090210(41) & 0.161(9) & 32.7 \\ 
f57 & 1.636219(26) & 18.93773(30) & 0.611164(9) & 0.159(9) & 32.3 \\ 
f58 & 2.031212(25) & 23.50940(29) & 0.492317(6) & 0.154(9) & 31.2 \\ 
f59 & 0.792700(27) & 9.17478(31) & 1.261509(43) & 0.151(9) & 30.6 \\ 
f60 & 4.125551(27) & 47.74944(31) & 0.2423918(15) & 0.147(9) & 29.9 \\ 
f61 & 2.683323(29) & 31.05698(34) & 0.3726722(41) & 0.134(9) & 27.1 \\ 
f62 & 4.019899(30) & 46.52662(35) & 0.2487623(19) & 0.130(9) & 26.4 \\ 
f63 & 3.148404(32) & 36.43987(37) & 0.3176211(32) & 0.126(9) & 25.6 \\ 
f64 & 3.143938(33) & 36.38818(38) & 0.3180723(34) & 0.125(9) & 25.3 \\ 
f65 & 3.546529(34) & 41.04779(39) & 0.2819658(27) & 0.117(9) & 23.7 \\ 
f66 & 3.440708(39) & 39.82301(45) & 0.2906378(32) & 0.103(9) & 20.9 \\ 
f67 & 3.520885(42) & 40.75099(49) & 0.2840194(34) & 0.093(9) & 18.8 \\ 
f68 & 2.910470(43) & 33.6860(5) & 0.3435866(51) & 0.090(9) & 18.2 \\ 
f69 & 3.860300(43) & 44.6794(5) & 0.2590474(29) & 0.089(9) & 18.1 \\ 
f70 & 3.455412(43) & 39.9932(5) & 0.2894009(39) & 0.086(9) & 17.5 \\ 
f71 & 3.473902(52) & 40.2072(6) & 0.2878607(41) & 0.080(9) & 16.3 \\ 
f72 & 4.65477(6) & 53.8747(7) & 0.2148333(27) & 0.077(9) & 15.7 \\ 
f73 & 3.281368(52) & 37.9788(6) & 0.3047510(47) & 0.076(9) & 15.5 \\ 
f74 & 3.012560(52) & 34.8676(6) & 0.331943(6) & 0.074(9) & 15.1 \\ 
f75 & 3.243810(52) & 37.5441(6) & 0.3082793(52) & 0.072(9) & 14.6 \\ 
f76 & 3.34979(6) & 38.7708(7) & 0.2985253(50) & 0.069(9) & 14.1 \\ 
f77 & 3.94288(6) & 45.6352(7) & 0.2536216(37) & 0.069(9) & 14.0 \\ 
f78 & 4.18767(6) & 48.4685(7) & 0.2387956(32) & 0.068(9) & 13.7 \\ 
f79 & 4.23285(6) & 48.9914(7) & 0.2362469(34) & 0.065(9) & 13.2 \\ 
f80 & 4.65285(7) & 53.8525(8) & 0.2149219(32) & 0.063(9) & 12.8 \\ 
f81 & 3.95572(7) & 45.7839(8) & 0.2527976(44) & 0.058(9) & 11.8 \\ 
f82 & 4.91806(7) & 56.9220(8) & 0.2033320(29) & 0.057(9) & 11.5 \\ 
f83 & 4.81245(8) & 55.6997(9) & 0.2077940(34) & 0.050(9) & 10.2 \\ 
f84 & 4.74814(10) & 54.9554(11) & 0.2106085(42) & 0.041(9) & 8.4 \\ 
\enddata
\end{deluxetable*}

\clearpage
\startlongtable
\begin{deluxetable*}{lllllc}
\tablenum{A4}
\tablecaption{Significant frequencies for KIC 5113357. Super-Nyquist real frequencies are flagged by asterisks. The first frequency in the table (f1) has the second-highest amplitude, and 3 of its harmonics, listed next, are found in the amplitude spectrum. This frequency may be a binary orbital frequency, and KIC 5113357 may be a contact eclipsing binary.\label{tab:A4}}
\tablewidth{0pt}
\tablehead{
\colhead{Frequency} & \colhead{Frequency} & \colhead{Frequency} & \colhead{Period} & \colhead{Amplitude} & \colhead{S/N } \\
\colhead{number} & \colhead{(c/d)} & \colhead{($\mu$Hz)} & \colhead{(days)} & \colhead{(ppt)} &
}
\startdata
f1 & 0.2475546(8) & 2.865216(9) & 4.039511(13) & 1.3662(34) & 396.4 \\ 
2 f1 & 0.4951094 & 5.730433 & 2.019756 & 0.0621(34) & 18.0 \\ 
3 f1 & 0.7426641 & 8.595649 & 1.346504 & 0.2178(34) & 63.2 \\ 
4 f1 & 0.9902188 & 11.460866 & 1.009878 & 0.0185(34) & 5.4 \\ 
\hline
f2 & 17.18742049(26) & 198.928478(3) & 0.0581820872(10) & 3.6451(34) & 1057.7 \\ 
f3 & 22.1956535(14) & 256.894138(16) & 0.0450538660(29) & 0.8488(34) & 246.3 \\ 
f4 & 15.1003578(15) & 174.772660(17) & 0.066223596(7) & 0.7952(34) & 230.7 \\ 
f5 & 16.4623012(18) & 190.535894(21) & 0.060744849(7) & 0.6863(34) & 199.1 \\ 
f6 & 17.0414043(19) & 197.238476(22) & 0.058680609(7) & 0.6453(34) & 187.2 \\ 
f7* & 30.7419943(24) & 355.810120(28) & 0.0325287938(25) & 0.5136(34) & 149.0 \\ 
f8* & 32.5167418(24) & 376.351179(28) & 0.0307533886(23) & 0.5114(34) & 148.4 \\ 
f9 & 18.0132124(25) & 208.486255(29) & 0.055514806(8) & 0.4887(34) & 141.8 \\ 
f10 & 15.9041145(25) & 184.075400(29) & 0.062876810(10) & 0.4836(34) & 140.3 \\ 
f11 & 15.0455407(27) & 174.138203(31) & 0.066464876(12) & 0.4828(34) & 140.1 \\ 
f12* & 31.2174329(31) & 361.312881(36) & 0.0320333834(31) & 0.3926(34) & 113.9 \\ 
f13 & 13.0846732(41) & 151.442977(47) & 0.076425293(24) & 0.3002(34) & 87.1 \\ 
f14* & 33.0103161(41) & 382.063844(47) & 0.0302935603(38) & 0.2973(34) & 86.3 \\ 
f15 & 19.0707264(43) & 220.72600(5) & 0.052436387(13) & 0.2715(34) & 78.8 \\ 
f16* & 26.8970276(43) & 311.30819(5) & 0.037178829(6) & 0.2646(34) & 76.8 \\ 
f17 & 15.6934765(52) & 181.63746(6) & 0.063720741(21) & 0.2469(34) & 71.6 \\ 
f18 & 16.504500(7) & 191.02431(8) & 0.060589534(27) & 0.1694(34) & 49.1 \\ 
f19 & 16.068174(8) & 185.97424(9) & 0.062234822(29) & 0.1662(34) & 48.2 \\ 
f20 & 15.047725(8) & 174.16349(9) & 0.066455225(36) & 0.1625(34) & 47.1 \\ 
f21 & 15.480492(9) & 179.17237(10) & 0.064597428(36) & 0.1408(34) & 40.9 \\ 
f22 & 15.264686(9) & 176.67461(10) & 0.065510680(38) & 0.1355(34) & 39.3 \\ 
f23 & 17.089691(10) & 197.79736(11) & 0.058514806(32) & 0.1293(34) & 37.5 \\ 
f24* & 27.376876(10) & 316.86200(12) & 0.036527175(14) & 0.1224(34) & 35.5 \\ 
f25 & 14.840393(10) & 171.76381(12) & 0.067383658(49) & 0.1152(34) & 33.4 \\ 
f26 & 16.165176(10) & 187.09695(12) & 0.061861372(41) & 0.1142(34) & 33.1 \\ 
f27 & 14.931307(11) & 172.81606(13) & 0.066973370(50) & 0.1091(34) & 31.6 \\ 
f28 & 23.675376(11) & 274.02056(13) & 0.042237975(21) & 0.1067(34) & 31.0 \\ 
f29 & 22.941774(11) & 265.52980(13) & 0.043588607(22) & 0.1064(34) & 30.9 \\ 
f30 & 15.723990(12) & 181.99063(14) & 0.063597087(49) & 0.1015(34) & 29.5 \\ 
f31 & 17.663605(13) & 204.43988(15) & 0.056613583(41) & 0.0974(34) & 28.3 \\ 
f32 & 17.292727(13) & 200.14731(15) & 0.057827777(43) & 0.0949(34) & 27.5 \\ 
f33 & 15.945179(14) & 184.55069(16) & 0.062714877(54) & 0.0899(34) & 26.1 \\ 
f34 & 21.973593(15) & 254.32400(17) & 0.045509168(30) & 0.0838(34) & 24.3 \\ 
f35 & 15.275913(15) & 176.80456(17) & 0.06546253(6) & 0.0827(34) & 24.0 \\ 
f36* & 27.099141(15) & 313.64747(17) & 0.036901538(21) & 0.0814(34) & 23.6 \\ 
f37 & 15.984584(16) & 185.00676(18) & 0.06256027(6) & 0.0806(34) & 23.4 \\ 
f38 & 12.480080(16) & 144.44538(18) & 0.08012768(10) & 0.0786(34) & 22.8 \\ 
f39 & 16.384703(17) & 189.63777(20) & 0.06103253(7) & 0.0703(34) & 20.4 \\ 
f40 & 20.511689(17) & 237.40381(20) & 0.048752689(42) & 0.0696(34) & 20.2 \\ 
f41* & 26.449871(18) & 306.13277(21) & 0.037807368(25) & 0.0683(34) & 19.8 \\ 
f42 & 20.301058(18) & 234.96596(21) & 0.049258515(44) & 0.0677(34) & 19.6 \\ 
f43* & 25.350242(18) & 293.40558(21) & 0.039447354(28) & 0.0675(34) & 19.6 \\ 
f44* & 25.963457(18) & 300.50298(21) & 0.038515671(27) & 0.0672(34) & 19.5 \\ 
f45 & 15.184358(18) & 175.74489(21) & 0.06585724(8) & 0.0670(34) & 19.4 \\ 
f46 & 17.778456(19) & 205.76917(22) & 0.05624785(6) & 0.0660(36) & 19.1 \\ 
f47 & 15.501081(19) & 179.41067(22) & 0.06451163(8) & 0.0646(34) & 18.7 \\ 
f48 & 15.628797(19) & 180.88886(22) & 0.06398444(8) & 0.0639(34) & 18.6 \\ 
f49* & 29.229992(19) & 338.31010(22) & 0.034211435(22) & 0.0638(34) & 18.5 \\ 
f50 & 15.840636(19) & 183.34070(22) & 0.06312877(8) & 0.0635(34) & 18.4 \\ 
f51 & 23.162114(21) & 268.08003(24) & 0.043173951(38) & 0.0598(34) & 17.4 \\ 
f52* & 31.157840(22) & 360.62315(25) & 0.032094650(22) & 0.0576(36) & 16.7 \\ 
f53 & 16.712736(22) & 193.43445(25) & 0.05983460(8) & 0.0571(34) & 16.6 \\ 
f54 & 21.920848(23) & 253.71353(27) & 0.045618671(47) & 0.0535(34) & 15.5 \\ 
f55* & 29.946360(24) & 346.60139(28) & 0.033393040(27) & 0.0513(34) & 14.9 \\ 
f56* & 30.339310(24) & 351.14943(28) & 0.032960538(27) & 0.0510(34) & 14.8 \\ 
f57 & 16.939838(24) & 196.06294(28) & 0.05903244(8) & 0.0507(34) & 14.7 \\ 
f58 & 1.692137(24) & 19.58492(28) & 0.590968(8) & 0.0500(34) & 14.5 \\ 
f59* & 27.591413(25) & 319.34506(29) & 0.036243160(32) & 0.0497(34) & 14.4 \\ 
f60 & 18.314286(25) & 211.97091(29) & 0.05460218(7) & 0.0485(34) & 14.1 \\ 
f61 & 21.802902(26) & 252.34841(30) & 0.045865453(54) & 0.0479(34) & 13.9 \\ 
f62 & 6.067855(27) & 70.22981(31) & 0.1648028(7) & 0.0460(34) & 13.4 \\ 
f63 & 21.247904(27) & 245.92482(31) & 0.04706346(6) & 0.0461(34) & 13.4 \\ 
f64* & 35.055539(27) & 405.73541(31) & 0.028526162(22) & 0.0455(34) & 13.2 \\ 
f65 & 14.959874(27) & 173.14669(31) & 0.06684548(13) & 0.0450(34) & 13.1 \\ 
f66 & 12.273925(28) & 142.05932(32) & 0.08147353(19) & 0.0448(34) & 13.0 \\ 
f67 & 20.805080(28) & 240.79954(32) & 0.04806518(6) & 0.0446(34) & 13.0 \\ 
f68 & 15.460037(28) & 178.93562(32) & 0.06468289(12) & 0.0444(34) & 12.9 \\ 
f69 & 15.399389(28) & 178.23368(32) & 0.06493763(12) & 0.0446(34) & 12.9 \\ 
f70* & 27.464285(29) & 317.87368(33) & 0.036410922(38) & 0.0425(34) & 12.3 \\ 
f71 & 17.538467(29) & 202.99152(34) & 0.05701752(9) & 0.0420(34) & 12.2 \\ 
f72 & 17.594314(30) & 203.63790(35) & 0.05683653(10) & 0.0400(34) & 11.6 \\ 
f73* & 26.069435(31) & 301.72958(36) & 0.038359097(45) & 0.0396(34) & 11.5 \\ 
f74 & 16.319956(31) & 188.88838(36) & 0.06127467(12) & 0.0394(34) & 11.4 \\ 
f75 & 21.284908(32) & 246.35311(37) & 0.04698164(7) & 0.0392(35) & 11.4 \\ 
f76 & 21.702645(32) & 251.18803(37) & 0.04607732(7) & 0.0394(35) & 11.4 \\ 
f77 & 1.440760(32) & 16.67547(37) & 0.694077(15) & 0.0387(34) & 11.2 \\ 
f78* & 28.690746(33) & 332.06883(38) & 0.034854442(39) & 0.0378(34) & 11.0 \\ 
f79 & 21.613439(33) & 250.15555(38) & 0.04626751(7) & 0.0371(34) & 10.8 \\ 
f80 & 18.085498(33) & 209.32290(38) & 0.05529291(10) & 0.0372(34) & 10.8 \\ 
f81 & 21.630309(34) & 250.35081(39) & 0.04623142(7) & 0.0364(34) & 10.6 \\ 
f82 & 21.282111(35) & 246.32074(40) & 0.04698781(8) & 0.0364(35) & 10.6 \\ 
f83 & 14.772660(37) & 170.97987(43) & 0.06769261(17) & 0.0328(34) & 9.5 \\ 
f84 & 16.202320(38) & 187.52686(44) & 0.06171954(15) & 0.0319(34) & 9.3 \\ 
f85 & 23.438314(39) & 271.27679(45) & 0.04266518(7) & 0.0315(34) & 9.1 \\ 
f86 & 19.723069(40) & 228.27627(46) & 0.05070204(10) & 0.0306(34) & 8.9 \\ 
f87* & 27.841771(40) & 322.24272(46) & 0.035917255(52) & 0.0305(34) & 8.9 \\ 
f88 & 21.700664(43) & 251.1651(5) & 0.04608153(9) & 0.0293(35) & 8.5 \\ 
f89* & 25.836843(41) & 299.03754(48) & 0.03870442(6) & 0.0292(34) & 8.5 \\ 
f90 & 19.809345(43) & 229.27483(50) & 0.05048122(10) & 0.0285(34) & 8.3 \\ 
f91* & 34.856205(43) & 403.4283(5) & 0.028689293(36) & 0.0282(34) & 8.2 \\ 
f92* & 24.669835(43) & 285.5305(5) & 0.04053533(7) & 0.0283(34) & 8.2 \\ 
f93 & 5.623207(43) & 65.08342(50) & 0.1778344(14) & 0.0284(34) & 8.2 \\ 
f94 & 18.578574(43) & 215.0298(5) & 0.05382545(13) & 0.0280(34) & 8.1 \\ 
f95 & 21.536072(43) & 249.2601(5) & 0.04643371(9) & 0.0274(34) & 8.0 \\ 
f96* & 27.416292(43) & 317.3182(5) & 0.03647466(6) & 0.0276(34) & 8.0 \\ 
f97* & 34.489964(43) & 399.1894(5) & 0.028993944(39) & 0.0258(34) & 7.5 \\ 
f98 & 14.733437(52) & 170.5259(6) & 0.06787283(22) & 0.0255(34) & 7.4 \\ 
f99 & 1.272084(52) & 14.7232(6) & 0.786111(30) & 0.0253(34) & 7.3 \\ 
f100* & 36.496310(52) & 422.4110(6) & 0.027400027(39) & 0.0250(37) & 7.2 \\ 
f101 & 23.193794(52) & 268.4467(6) & 0.04311497(9) & 0.0248(34) & 7.2 \\ 
f102* & 25.096668(52) & 290.4707(6) & 0.03984592(8) & 0.0238(34) & 6.9 \\ 
f103 & 8.025471(52) & 92.8874(6) & 0.1246033(8) & 0.0238(34) & 6.9 \\ 
f104 & 5.787115(52) & 66.9805(6) & 0.1727976(16) & 0.0232(35) & 6.7 \\ 
f105 & 6.927353(52) & 80.1777(6) & 0.1443552(12) & 0.0232(34) & 6.7 \\ 
f106 & 21.713676(52) & 251.3157(6) & 0.04605393(12) & 0.0228(34) & 6.6 \\ 
f107* & 35.349730(52) & 409.1404(6) & 0.028288758(44) & 0.0223(34) & 6.5 \\ 
f108 & 13.554527(52) & 156.8811(6) & 0.07377608(30) & 0.0222(34) & 6.5 \\ 
f109* & 25.75142(6) & 298.0489(7) & 0.03883280(8) & 0.0220(34) & 6.4 \\ 
f110 & 5.78434(6) & 66.9484(7) & 0.1728804(17) & 0.0217(35) & 6.3 \\ 
f111 & 21.27187(6) & 246.2023(7) & 0.04701041(13) & 0.0213(34) & 6.2 \\ 
f112 & 5.00818(6) & 57.9651(7) & 0.1996732(23) & 0.0214(34) & 6.2 \\ 
f113 & 10.84563(6) & 125.5282(7) & 0.09220299(50) & 0.0211(34) & 6.1 \\ 
f114 & 19.96522(6) & 231.0790(7) & 0.05008709(15) & 0.0212(34) & 6.1 \\ 
f115 & 20.95324(6) & 242.5144(7) & 0.04772530(13) & 0.0210(34) & 6.1 \\ 
f116* & 26.29695(6) & 304.3629(7) & 0.03802722(9) & 0.0198(34) & 5.8 \\ 
f117 & 23.90478(6) & 276.6758(7) & 0.04183262(10) & 0.0201(34) & 5.8 \\ 
f118 & 13.39652(6) & 155.0524(7) & 0.07464622(35) & 0.0195(34) & 5.7 \\ 
f119 & 11.56630(6) & 133.8693(7) & 0.08645800(46) & 0.0197(34) & 5.7 \\ 
f120* & 27.18498(6) & 314.6410(7) & 0.03678502(8) & 0.0193(34) & 5.6 \\ 
f121* & 26.52131(7) & 306.9597(8) & 0.03770552(9) & 0.0184(34) & 5.3 \\ 
\enddata
\end{deluxetable*}

\clearpage
\startlongtable
\begin{deluxetable*}{lcccccc}
\tablenum{A5}
\tablecaption{Significant frequencies for KIC 5112843. The spectrum shows only harmonics of the two lowest frequencies.  The average background noise level in the residual after pre-whitening these frequencies is 0.0055 ppt, with higher noise level at lower frequencies.\label{tab:A5}}
\tablewidth{0pt}
\tablehead{
\colhead{Frequency} & \colhead{Frequency} & \colhead{Frequency} & \colhead{Period} &
\colhead{Amplitude} & \colhead{S/N } \\
\colhead{number} & \colhead{(c/d)} & \colhead{($\mu$Hz)} & \colhead{(days)} &
\colhead{(ppt)} &
}
\startdata
f1 & 2.60356230(20) & 30.1338229(23) & 0.384089138(30) & 0.927(18) & 168.55 \\
2 f1 & 5.2071245 & 60.2676447 & 0.1920446 & 13.058(18) & 2374.18\\ 
3 f1 & 7.8106868 & 90.4014676 & 0.1280297 & 0.208(18) & 37.82 \\
4 f1 & 10.414249 & 120.5352894 & 0.0960223 & 0.087(18) & 15.82 \\
8 f1 & 20.8284981 & 241.0705799 & 0.0480111 & 0.023(18) & 4.18 \\
\hline
f2 & 2.86782760(10) & 33.1924491(12) & 0.348695600(12) & 0.707(18) & 128.55 \\
2 f2 & 5.7356552 & 66.3848982 & 0.1743480 & 21.344(18) & 3880.73 \\ 
3 f3 & 8.6034828 & 99.5773472 & 0.1162320 & 0.616(18) & 112.0 \\
4 f2 & 11.4713104 & 132.7697963 & 0.0871740 & 5.432(18) & 987.64 \\
5 f2 & 17.2069657 & 199.1546956 & 0.0581160 & 1.887(18) & 343.09 \\
6 f2 & 22.9426209 & 265.5395938 & 0.0435870 & 0.394(18) & 71.64 \\
7 f2 & 14.3391381 & 165.9622465 & 0.0697392 & 0.243(18) & 44.18 \\
8 f2 & 28.6782761 & 331.9244919 & 0.0348696 & 0.103(18) & 18.73 \\
9 f2 & 20.0747933 & 232.347145 & 0.0498137 & 0.067(18) & 12.18 \\        
10 f2 & 25.8104485 & 298.732043 & 0.0387440 & 0.031(18) & 5.64 \\
\enddata
\end{deluxetable*}

\end{document}